\documentclass[twocolumn,aps,prd,floats,eqsecnum,preprintnumbers,showpacs]{revtex4}

\usepackage{latexsym}
\usepackage{epsfig}
\usepackage{amssymb}
\usepackage{amsfonts}
\usepackage{graphicx}

\def\laq{\raise 0.4ex\hbox{$<$}\kern -0.8em\lower 0.62ex\hbox{$\sim$}}
\def\gaq{\raise 0.4ex\hbox{$>$}\kern -0.7em\lower 0.62ex\hbox{$\sim$}}

\font\tenbb=msbm10
\font\sevenbb=msbm7
\font\fivebb=msbm5
\newfam\bbfam
\textfont\bbfam=\tenbb \scriptfont\bbfam=\sevenbb
\scriptscriptfont\bbfam=\fivebb

\newcommand{\beeq}{\begin{equation}}
\newcommand{\eneq}{\end{equation}}
\newcommand{\beeqa}{\begin{eqnarray}}
\newcommand{\eneqa}{\end{eqnarray}}

\begin{document}
\preprint{TRINLAT-02/09, UPRF-2003-07}

\title{\large\bf The Supersymmetric Ward-Takahashi Identity in 1-Loop \\
Lattice Perturbation Theory: General Procedure}
\author{Alessandra Feo }
\affiliation{{\it School of Mathematics, Trinity College, Dublin 2, Ireland \\ and \\
           Dipartimento di Fisica, Universit\'a di Parma and INFN Gruppo \\ 
             Collegato di Parma, Parco Area delle Scienze, 7/A, 43100 Parma, Italy}}

\begin{abstract}
The one-loop corrections to the lattice supersymmetric Ward-Takahashi
identity (WTi) are investigated in the off-shell regime.  In the
Wilson formulation of the $N=1$ supersymmetric Yang-Mills (SYM)
theory, supersymmetry (SUSY) is broken by the lattice, by the Wilson
term and is softly broken by the presence of the gluino mass.
However, the renormalization of the supercurrent can be realized in a
scheme that restores the continuum supersymmetric WTi (once the
on-shell condition is imposed).  The general procedure used to
calculate the renormalization constants and mixing coefficients for
the local supercurrent is presented. The supercurrent not only mixes
with the gauge invariant operator $T_\mu$. An extra mixing with other
operators coming from the WTi appears.  This extra mixing survives in
the continuum limit in the off-shell regime and cancels out when the
on-shell condition is imposed and the renormalized gluino mass is set
to zero.  Comparison with numerical results are also presented.
\end{abstract}
\maketitle

\section{Introduction}
\label{sec1}

SUSY or fermion-boson symmetry is one of the most exciting topics in field theory.
From a theoretical point of view SUSY plays a fundamental role in string theory.
There are many strong phenomenological motivations for believing that SUSY is realized in Nature 
in a spontaneoulsy broken form. The SUSY breaking mechanisms are requested in order to
produce a low energy 4-dimensional effective action with a residual $N=1$ SUSY.
For the other hand, non-perturbative studies of supersymmetric gauge theories turn out to have remarkably rich 
properties which are of great physical interest, as has been pointed out in \cite{seiberg}.
For this reason, much effort has been dedicated to formulating a lattice version of 
supersymmetric theories (for a recent review in SUSY on the lattice with a complete
list of relevant references, see \cite{feo5}).
More recently, related interesting results in SUSY can be found in 
\cite{hiller,rey,catterall,beccaria,campostrini,campostrini2,itoh,kaplan,kaplan2,giedt}.
Some of these formulations try to realize chiral gauge theories on the lattice
with an exact chiral gauge symmetry \cite{luscher1,luscher2,luscher3,luscher4}.
The lattice formalism is a powerful tool to extract non-perturbative
dynamics of field theories and may be able to provide additional information and 
confirm or improve theoretical expectations.

To formulate SUSY on the lattice we follow the ideas of Curci and Veneziano \cite{curci}. 
They propose to give up manifest SUSY on the lattice, and instead, to restore it in the continuum 
limit. 
In \cite{curci}, the Wilson formulation for the $N=1$ SYM theory, which is the simplest 
SUSY gauge theory and corresponds to the SUSY gluodynamics, is adopted. 
For SU$(N_c)$ it has $(N_c^2 - 1)$ gluons and the same number of massless Majorana fermions
(gluinos), in the adjoint representation of the color group.

SUSY is broken explicitly by the Wilson term and the finite lattice spacing. 
In addition, a soft breaking due to the introduction of the gluino mass is present.
In \cite{curci} it is proposed that SUSY can be recovered in the continuum 
limit by tuning the bare gauge coupling, $g_0$, and the gluino mass, $m_{\tilde{g_0}}$, 
to the SUSY point, $m_{\tilde{g_0}}=0$, which also coincides with the chiral point.
In \cite{montvay,pap,kirchner,kirchner2,campos,farchioni,montvay3}, the DESY-M\"unster-Roma Collaboration
have investigated these issues for the $SU(2)$ gauge group (some first results have
been obtained for $SU(3)$ \cite{feo2}), simulating the theory with 
a dynamical gluino using the multi-bosonic algorithm \cite{luscher5} with 
a two-step variant called the (TSMB) algorithm \cite{montvay2}
(while quenched results are in \cite{vladikas}). 

Another independent way to study the SUSY (chiral) limit in the Wilson formulation of Curci and Veneziano, 
is through the study of the SUSY WTi. On the lattice, it contain explicit SUSY breaking terms and 
the SUSY limit is defined to be the point in the parameter space where these breaking terms 
vanish and the SUSY WTi recovers its continuum form. 
These issues have been investigated numerically in \cite{farchioni} in the on-shell regime.

In this paper, the general procedure used to determine the renormalization constants 
and mixing coefficients for the local definition of the supercurrent,
in the off-shell regime, is explained.
It is shown that, when the operator insertion involves elementary fields, 
the supercurrent not only mixes with the gauge invariant operator $T_\mu$,
as have been claimed in \cite{curci}. The supercurrent contains also 
non-Lorentz covariant terms which survive in the continuum, in the off-shell regime. 
These non-Lorentz breaking terms cancel out when the on-shell condition on the gluino is imposed and
the continuum SUSY WTi is recovered.
Preliminary studies have been presented in \cite{feo,feo4}.

The paper is organized as follows. In Sec.~\ref{sec2} the Curci and Veneziano lattice formulation of 
the $N=1$ SYM theory is presented, together with the lattice action and the vertices used for 
the calculation. 
In Sec.~\ref{sec3} the SUSY WTi on the lattice are written and the renormalization
procedure explained. 
The calculation of the renormalization constant for the supercurrent is presented in Sec.~\ref{sec4}. 
Discussions and outlook are summarized in Sec.~\ref{sec5}. 
In Appendices~\ref{appendixPERT},~\ref{appendixVERTICES} and~\ref{appendixOFFSHELL},
some details of the calculation are showed. 

\section{Lattice Formulation}
\label{sec2}

In the Wilson formulation of the $N=1$ SYM theory \cite{curci}, the gluonic 
part of the action is the standard plaquette one
\beeq
S_g = \frac{\beta}{2} \sum_{x} \sum_{\mu \nu} \bigg(1 - \frac{1}{N_c} 
\mbox{Re} \, \mbox{Tr} \, P_{\mu \nu}(x) \bigg) \, ,
\label{a11}
\eneq
\noindent
where the plaquette operator is defined as \cite{muenster}
\beeq
P_{\mu \nu}(x) = U^\dagger_\nu(x) U^\dagger_\mu(x+\hat{\nu}) U_\nu(x+\hat{\mu}) U_\mu(x)\, , 
\label{a12}
\eneq
\noindent
and the bare coupling is given by $\beta \equiv 2 N_c/g_0^2$.
For Wilson fermions the fermionic part of the action reads
\beeqa
&& S_f = \sum_x a^4 \mbox{Tr} \bigg[ \nonumber \\ 
&& \frac{1}{2a} \Bigg( \bar\lambda(x)( \gamma_\mu - r) U_\mu^\dagger(x)
\lambda(x + a \hat\mu) U_\mu(x) - \nonumber \\ 
&&  \bar\lambda(x + a \hat\mu) (\gamma_\mu + r) U_\mu(x)
\lambda(x) U_\mu^\dagger(x) \Bigg) + \nonumber  \\
&&  \bigg( m_0 + \frac{4r}{a} \bigg) \bar\lambda(x) \lambda(x) \bigg] \, ,
\label{a13}
\eneqa
where $m_0$ is the gluino bare mass and $a$ is the lattice spacing.
The fermionic field (gluino), $\lambda(x) = \lambda^a(x) T^a $, is a Majorana spinor 
in the adjoint representation of the gauge group.
The symbol $\mbox{Tr}$ implies the trace over the color indices. The normalization
is given by $\mbox{Tr}(T^a T^b) = \frac{1}{2} \delta_{ab}$. 
In this paper, only the case $N_c=2$ is considered, for which the adjoint gluino field 
is expressed in terms of Pauli matrices $\sigma_k$ as
\beeq
\lambda = \sum_{k=1}^3 \frac{1}{2} \sigma_k \lambda^k \, .
\eneq
The gluino field $\lambda(x)$ satisfies the Majorana condition
\beeq
\lambda(x) = C \bar \lambda^T(x) \, ,
\eneq
where $C = \gamma_2 \gamma_0$, is the charge conjugation operator.
Our matrix convention for the Euclidean $\gamma$ matrices is as follow,
\beeq
\gamma_0 = \left(
           \begin{array}{cc}
             0 & 1 \\
             1 & 0 
           \end{array}
           \right) 
\eneq
and 
\beeq
\gamma_k = \left(
           \begin{array}{cc}
             0 & - i \sigma^k \\
             i \sigma^i & 0 
           \end{array}
           \right) \, .
\eneq
The matrix $\gamma_5$ is defined to be
\beeq
\gamma_5 \equiv \gamma_1 \gamma_2 \gamma_3 \gamma_0  = \left(
             \begin{array}{cc}
             1 & 0 \\
             0 & -1 
           \end{array}
           \right) 
\eneq
and the matrix $\sigma_{\mu \nu}$ is
\beeq
\sigma_{\mu \nu} = \frac{i}{2} [\gamma_\mu , \gamma_\nu] \, .
\eneq
The anti-commutator property is 
\beeq
\big\{ \gamma_\mu, \gamma_\nu \big\} = 2 \delta_{\mu \nu} \, . 
\eneq
Finally, the Wilson parameter is fixed to be $r=1$.

SUSY is not realized on the lattice because, as the Poincar\'e algebra, a sector of the superalgebra, 
is lost. SUSY is explicitly broken in the action
(\ref{a11},\ref{a13}) by the lattice itself, by the gluino mass term and 
by the Wilson term.
Nevertheless, one can still define some transformations that reduce to the continuum
supersymmetric ones, in the limit $a \to 0$. One choice is \cite{taniguchi,galla} 
\footnote{Our definition of the link variable $U_\mu(x)$ differs from that of \cite{taniguchi}
(see our definition of the plaquette \ref{a12}); the two definitions are related by 
Hermitian conjugation.}:
\beeqa
\delta U_\mu(x) &=& -a g_0 U_\mu(x) \bar \xi(x) \gamma_\mu \lambda(x) - \nonumber \\ 
&& a g_0 \bar \xi(x) \gamma_\mu \lambda(x + a \hat\mu) U_\mu(x) \, , \nonumber \\
\delta U_\mu^\dagger(x)& =&
a g_0 \bar \xi(x) \gamma_\mu \lambda(x) U_\mu^\dagger(x) + \nonumber \\ 
&& a g_0 U_\mu^\dagger(x) \bar \xi(x) \gamma_\mu \lambda(x + a \hat\mu) \, , \nonumber \\
\delta \lambda(x) & = &
- \frac{i}{g_0} \sigma_{\rho\tau} {\cal G}_{\rho \tau}(x) \xi(x) \, , \nonumber \\
\delta \bar \lambda(x) & =&
 \frac{i}{g_0} \bar \xi(x) \sigma_{\rho\tau} {\cal G}_{\rho \tau}(x) \, ,
\label{a3}
\eneqa
where $\xi(x)$ and $\bar \xi(x)$ are infinitesimal Majorana fermionic parameters, while 
${\cal G}_{\rho \tau}(x)$ is the clover plaquette operator,
\beeqa
{\cal G}_{\rho \tau}(x) & = & - \frac{1}{8 a^2}
\bigg( P_{\rho \tau}(x) - P_{\tau \rho}(x) + \nonumber \\ 
&& P_{-\rho, -\tau}(x) - P_{-\tau, -\rho}(x) + P_{\tau, -\rho}(x) - \nonumber \\ 
&& P_{-\rho, \tau}(x) + P_{-\tau, \rho}(x) - P_{\rho, -\tau}(x) \bigg) \, .
\eneqa
\noindent
Weak coupling perturbation theory is developed by writing the link variable as 
\beeq
U_\mu(x) = e^{-a A_\mu(x)} \, , 
\eneq
and expanding it in terms of $g_0$. Here the 
gluon field is defined to be $A_\mu(x) = -i g_0 A_\mu^b(x) T^b$.

In order to calculate the 1-loop corrections to the SUSY WTi (which correspond to $O(g_0^2)$), 
we need two kinds of gluon-gluino interaction vertices. 
The gluon-gluino vertex,
\beeqa
V_{1 \mu}^{ab,c}(p,q) & = & g_0 f^{abc} \bigg[ \gamma_\mu \cos (\frac{p_\mu a}{2} + \frac{q_\mu a}{2}) - \nonumber \\ 
&& i r \sin(\frac{p_\mu a}{2} + \frac{q_\mu a}{2}) \bigg] \, , 
\eneqa
two-gluons-one-gluino vertex,
\beeqa
V_{2 \mu \nu}^{ab,cd}(k,p)& =& \frac{1}{2} a g_0^2 (f^{ace} f^{ebd} + f^{ade} f^{ebc}) \bigg[ \nonumber \\ 
&&  \hspace{-2.8 cm} i \gamma_\mu \sin (\frac{p_\mu a}{2} + \frac{q_\mu a}{2}) - r \cos (\frac{p_\mu a}{2} + \frac{q_\mu a}{2}) 
 \bigg] \delta_{\mu \nu} 
\eneqa
and the three-gluons vertex
\beeqa
G_{3 \mu \nu \lambda}^{abc}(k,k_1,k_2) &=& i g_0 f^{abc} \bigg[ \nonumber \\ 
&& \hspace{-2.0 cm} \delta_{\nu \lambda} \cos(\frac{k_\nu a}{2}) \sin (\frac{k_{2 \mu} a}{2} - \frac{k_{1 \mu} a}{2}) + \nonumber \\ 
&& \hspace{-2.0 cm} \delta_{\mu \lambda} \cos(\frac{k_{1 \lambda} a}{2}) \sin (\frac{k_\nu a}{2} - \frac{k_{2 \nu} a}{2}) + \nonumber \\ 
&& \hspace{-2.0 cm} \delta_{\mu \nu} \cos(\frac{k_{2\mu} a}{2}) \sin (\frac{k_{1 \lambda} a}{2} - \frac{k_\lambda a}{2}) \bigg] \, .
\eneqa
\noindent
These vertices are similar to QCD and the only difference is that the fermion is a 
Majorana fermion in the adjoint representation of the gauge group instead of the fundalmental one.

\section{SUSY WTi on the lattice} 
\label{sec3}

The vacuum expectation value of an operator ${\cal O}$ is defined to be
\beeq
\big< {\cal O} \big> = \int dU \, d\lambda \, {\cal O} \, e^{-S_{total}} \, , 
\eneq
where $S_{total}$ is the total action on the lattice. By applying an infinitesimal local
supersymmetric transformation, with a localized transformation parameter 
$\xi(x) $, the lattice WTi is written as \cite{feo},
\beeqa
&& \big< {\cal O} \nabla_\mu S_\mu(x) \big> -
2 m_0 \big< {\cal O} \chi(x) \big> + \bigg< \left. \frac{\delta{\cal O}}
{\delta \bar \xi(x)}\right|_{\xi = 0}\bigg> \nonumber \\ 
&& - \bigg< \left. {\cal O} \, \frac{\delta S_{GF}}{\delta \bar \xi(x)} 
\right|_{\xi = 0} \bigg> - \bigg< \left. {\cal O} \, \frac{\delta S_{FP}}
{\delta \bar \xi(x)}\right|_{\xi = 0}\bigg> = \nonumber \\ 
&& \hspace{1.0 cm} \big< {\cal O} X_S(x) \big> \, ,
\label{a2}
\eneqa
where $S_{GF}$ is the gauge fixing term, $S_{FP}$ is the Faddeev-Popov term and 
$ \frac{\delta{\cal O}}{\delta \bar \xi(x)}|_{\xi = 0}$ represents the contact terms
(see Appendices~\ref{appendixPERT} and~\ref{appendixVERTICES} for definitions).
This WTi is also discussed in \cite{dewit}.
$X_S(x)$ is the symmetry breaking term coming from the fact that the action 
is not fully invariant under (\ref{a3}).
Usually $X_S(x)$ is a complicated function of the link variables and the fermionic variables
\cite{taniguchi}, and its specific form depends on the choice of the lattice supercurrent. 

Let us define the lattice local supercurrent as
\beeq
S_\mu(x) = -\frac{2 i}{g_0} \, \mbox{Tr} \, \bigg\{ {\cal G}_{\rho \tau}(x)
           \sigma_{\rho \tau} \gamma_\mu \lambda(x) \bigg\} \, ,
\label{a15}
\eneq
while $\nabla_\mu$ is the symmetric lattice derivative
\beeq
\nabla_\mu f(x) = \frac{1}{2 a} (f(x + a \hat{\mu}) - f(x - a \hat{\mu})) 
\eneq
\noindent
and $\chi(x)$ corresponds to the gluino mass term
\beeq
\chi(x) = \frac{i}{g_0} \, \mbox{Tr} \, \bigg\{ {\cal G}_{\rho \tau}(x) \sigma_{\rho \tau} 
\lambda(x) \bigg\} \, .
\label{a30}
\eneq

\noindent
In order to renormalize the lattice WTi the operator mixing has to be 
taken into account. The standard way to renormalize the supercurrent is to define a substracted 
$\bar X_S$, whose expectation value is forced to vanish in the limit $a \to 0$ \cite{bochicchio,testa}.
In the case in which the operator insertion ${\cal O}$ in Eq.~(\ref{a2}) is gauge invariant,
$X_S$ mixes with the following operators of equal or lower dimension \cite{vladikas}
\beeqa
X_S(x)& =&  \bar{X}_S(x) - (Z_S - 1) \nabla_\mu S_\mu(x) \nonumber \\ 
&& - 2 \tilde{m} \chi(x) - Z_T \nabla_\mu T_\mu(x) \, ,
\label{a1}
\eneqa
where the current $T_\mu$ reads
\beeq
T_\mu(x) = -\frac{2}{g} \, \mbox{Tr} \, \bigg\{ {\cal G}_{\mu \nu}(x)
               \gamma_\nu \lambda(x) \bigg\} \, .
\label{a33}
\eneq

On the other hand, if the operator insertion ${\cal O}$ is non-gauge invariant 
(i.e., the one involving elementary fiels),
the gauge dependence implies that operator mixing with non-gauge invariant terms has 
to be taken into account in the renormalization procedure. In this case Eq.~(\ref{a1})
is modified as \cite{feo,vladikas2}
\beeqa
\hspace{-2 cm} X_S(x) &=& \bar{X}_S(x) - (Z_S - 1) \nabla_\mu S_\mu(x) \nonumber \\ 
&& \hspace{-1 cm} - 2 \tilde{m} \chi(x) - Z_T \nabla_\mu T_\mu(x) - \sum_j Z_{B_j} B_j \, . 
\label{a45}
\eneqa
The $B_j$'s denote the occurence of mixing, not only with non-gauge invariant operators 
but also mixing with gauge invariant operators which do not vanish in the off-shell regime 
(but vanish in the on-shell regime). Consider, for example, the gauge invariant operator 
\beeq
B_0 = \frac{2}{g}\mbox{Tr}\bigg\{ \gamma_\rho (D_\tau {\cal G}_{\rho \tau}(x)) \lambda(x) \bigg\} \, ,
\label{a58}
\eneq
which is zero imposing the equations of motion (thus, is not considered in \cite{farchioni}), 
but in the off-shell regime is non-zero and must be considered \cite{vladikas3}. 
Other non-gauge invariant operators, which should be included in $B_j$ are
\beeq
B_1 = \frac{2}{g} \partial_\rho A_\rho \not \partial \lambda \, , \,
B_2 = \frac{2}{g} A_\rho \partial_\rho \not \partial \lambda \, , \, 
B_3 = \frac{2}{g} \not A \partial_\rho \partial_\rho \lambda \, ,
\label{a59}
\eneq
(also reported in \cite{taniguchi}).
Finally, non-Lorentz covariant terms coming from $\nabla_\mu S_\mu$, the gauge fixing term and contact terms, which 
appear in the off-shell regime, should also be taken in consideration.
Because the $B_j$ do not appear in the tree-level WTi, $Z_{B_j}$ should be $O(g^2)$ \cite{feo}.

Substituting (\ref{a45}) in (\ref{a2}) we obtain the renormalized WTi
%\begin{widetext}
\beeqa
&& Z_S \big< {\cal O} \nabla_\mu S_\mu(x) \big> + Z_T \big< {\cal O} \nabla_\mu T_\mu(x) \big> - \nonumber \\ 
&& 2 (m_0 - \tilde{m}) Z^{-1}_{\chi} \big< {\cal O} \chi^R(x) \big> + 
Z_{CT} \bigg< \frac{\delta {\cal O}} {\delta \bar \xi(x)}|_{\xi = 0} \bigg> - \nonumber \\ 
&& Z_{GF} \bigg< {\cal O} \, \frac{\delta S_{GF}}{\delta \bar \xi(x)}|_{\xi = 0} \bigg> -
Z_{FP} \bigg< {\cal O} \, \frac{\delta S_{FP}}{\delta \bar \xi(x)}|_{\xi = 0} \bigg> + \nonumber \\ 
&& \sum_j Z_{B_j} \big< {\cal O} B_j \big>=0 \, .
\label{a8}
\eneqa
%\end{widetext}
The contact terms, Faddeev-Popov term and gauge fixing term should 
be renormalized, that is why in Eq.~(\ref{a8})
the renormalization constants $Z_{CT}$, $Z_{GF}$ and $Z_{FP}$ are introduced.
$\big< {\cal O} B_j \big> $ can in principle do mixing with $S_\mu$ and $T_\mu$ \cite{feo}. 
This implies that $S_\mu$ not only mixes with $T_\mu$ as was predicted in \cite{curci}, but 
extra mixing with gauge variant operators and/or gauge invariant operators, 
which do not vanish in the off-shell regime, can appear. 
These extra mixing vanish by setting the renormalized gluino mass to zero 
and by imposing the on-shell condition on the gluino.

In the continuum, the existence of a renormalized SUSY WTi
\beeq
\partial_\mu S^R_\mu = 2 m_R Z_\chi \chi \, , 
\label{a48}
\eneq
is generally assumed, where $S_R$ is the renormalized supercurrent and 
$m_R$ is the renormalized gluino mass. 
For $m_R = 0$, we have SUSY while a non-vanishing value of $m_R$ breaks SUSY softly. 
It is generally assumed that SUSY is not anomalous (Eq.~(\ref{a48}) holds) and only the 
mass term is responsible for a soft breaking. However, in \cite{shamir} the question of 
whether non-perturbative effects may cause a SUSY anomaly has been raised.

It is tempting to associate the normalized continuum supercurrent as
\beeq
S_\mu^R = Z_S S_\mu + Z_T T_\mu \, ,
\label{a50}
\eneq 
in analogy with the lattice chiral WTi in QCD. This analogy fails, as has been pointed out in
\cite{testa}. Explicit 1-loop calculation may shed some light on this issue. If the correctly 
normalized supercurrent coincides with (\ref{a50}), then it is conserved when $m_R =0$. 
This is the restoration of SUSY in the continuum limit \cite{farchioni}. 

By using general renormalization group arguments (see for example, \cite{testa}), one can show that
$Z_S$, $Z_T$ and $Z_\chi$, being power substraction coefficients, do not depend on the 
renormalization scale $\mu$, defining the renormalization operator in Eq.~(\ref{a1}).
This imply that $Z_S = Z_S(g_0,m_0 a)$, $Z_T = Z_T(g_0,m_0 a)$ and $Z_\chi = Z_\chi(g_0,m_0 a)$.

In this paper, we are interesting to calculate the renormalization constant for 
the local supercurrent (\ref{a8}) and compare with Monte Carlo 
results in \cite{farchioni}. Notice that the relation between the 1-loop perturbative calculation
and the numerical one is $Z_T Z_S^{-1} \equiv Z_T|_{1-loop}$. This is because, 
$Z_S = 1 + O(g_0^2)$, while $Z_T = O(g_0^2)$. So it is enough to
calculate the coefficient $Z_T$ in 1-loop lattice perturbation theory (LPT). 
The numerical estimates are \cite{farchioni} $Z_T Z_S^{-1} = -0.039(7)$ for the point-split current 
and $Z_T Z_S^{-1} = 0.185(7)$ for the local current, both at $\beta = 2.3$.
An estimate of $Z_T Z_S^{-1}$ for the point-split current at $\beta = 2.3$ can be obtained 
from the 1-loop perturbative calculation in \cite{taniguchi}. At order $g_0^2$ the value   
is $Z_T|_{1-loop} = -0.074$ \cite{taniguchi}.
In this paper, the calculation of $Z_T|_{1-loop}$ for the local supercurrent is presented. 

In principle, each matrix element in Eq.~(\ref{a8}) is proportional to each element 
of the $\Gamma$-matrix base
\beeq
\Gamma = \left\{ 1, \gamma_5, \gamma_\alpha, \gamma_5 \gamma_\alpha, \sigma_{\alpha \rho} \right\} \, ,
\label{a31}
\eneq
but in order to determine $Z_T$, it is enough to calculate in Eq.~(\ref{a8}) the projections 
over two elements of the base (\ref{a31}). 

\section{Renormalization Constants} 
\label{sec4}
\noindent
We are now considering 
each matrix element in Eq.(\ref{a8}) with ${\cal O}$ (a non-gauge invariant operator) given by 
\beeq
{\cal O} := A_\nu^b(y)\, \bar \lambda^a(z) \, , 
\label{a14}
\eneq
In Fourier transformation (FT) we choose $p$ as the outcoming momemtun for the gluon field $A_\mu$ and $q$ 
the incoming momentum for the fermion field $\lambda$ (see Fig.~\ref{Fig1}). 
Each matrix element can be written as
\beeq
\big< A_\nu^b(y) \, \bar \lambda^a(z) \, C(x) \big> \stackrel{FT}{\Longrightarrow}
D_F(q) \cdot (C(p,q))_{amp} \cdot D_B(p) \cdot \delta_{ab}  \, ,
\label{a32}
\eneq
where $(C(p,q))_{amp}$ can be, i.e., $\nabla_\mu S_\mu$, $\nabla_\mu T_\mu$, etc.,  
with the external propagators amputated,
$D_F(q)$ and $D_B(p)$ are the full fermion and gluon propagators, 
respectively, while $\delta_{ab}$ is the color structure, similar to all diagrams.
The non-trivial part of the calculation is the determination of $(C(p,q))_{amp}$ for each matrix 
element in Eq.~(\ref{a8}). $\big< {\cal O} \chi(x) \big>$, is not considered as we set the renormalized 
gluino mass to zero.

In order to determine $Z_T$ one should pick up
from each matrix element of Eq.~(\ref{a8}) those terms which contains the same Lorentz structure 
as $S_\mu$ and $T_\mu$, to tree-level. Those operators which do not contain the same 
tree-level Lorentz structure than $S_\mu$ and $T_\mu$ do not enter in the determination of $Z_T$.
Below, we present the tree-level values of the different operator of Eq.~(\ref{a8}).
The calculation is straighforward. 

For the case of the supercurrent (\ref{a15}), the tree-level part reads 
\beeq
S_\mu^{(0)}(x) = - \frac{2 i}{g} \, \mbox{Tr} \, \bigg\{ (\partial_\rho A_\tau(x) -  
               \partial_\tau A_\rho(x)) \sigma_{\rho \tau} \gamma_\mu \lambda(x) \bigg\} \, .
\eneq
\noindent
Using $\mbox{Tr}(T^a T^b)=\frac{1}{2} \delta_{ab}$ for the traces and the antisymmetry of 
$\sigma_{\rho \tau}$ this expression becomes 
\beeq
S_\mu^{(0)}(x) = -2 \delta_{ab} \partial_\rho A_\tau^b(x) \sigma_{\rho \tau} \gamma_\mu 
\lambda^a(x) \, . 
\eneq
\noindent
or in FT,
\beeqa
\tilde{S}_\mu^{(0)}(r) &=& \int d^4x e^{i r \cdot x} S_\mu^{(0)}(x)  \nonumber \\
\hspace{-1cm}  &=& -2 i \delta_{ab} \sigma_{\rho \tau} \gamma_\mu \int d^4x \int \frac{d^4p}{(2 \pi)^4} 
\int \frac{d^4q}{(2 \pi)^4} \times \nonumber \\ 
&& \times e^{i(r-p+q) x} p_\rho \tilde A_\tau^b(p) \tilde \lambda^a(q) \nonumber \\
\hspace{-1cm}  &=& -2 i \delta_{ab} \sigma_{\rho \tau} \gamma_\mu \int \frac{d^4p}{(2 \pi)^4} 
\int \frac{d^4q}{(2 \pi)^4} \times \nonumber \\ 
&& \times \delta(r-p+q) p_\rho \tilde A_\tau^b(p) \tilde \lambda^a(q) \, ,
\label{a53}
\eneqa
so we can define the vertex 
\beeq
\tilde{S}^{ab}_{\mu, \tau}(p,q) = -2 i \delta_{ab} \sigma_{\rho \tau} \gamma_\mu p_\rho \, .
\label{a54}
\eneq

Concerning the operator $T_\mu$ in Eq.~(\ref{a33}), the tree-level is 
\beeq
T_\mu^{(0)}(x) = i \delta_{ab} (\partial_\mu A_\tau^b(x) - \partial_\tau A_\mu^b(x)) \gamma_\tau \lambda^a(x)
\eneq
after the FT, we define the corresponding tree-level vertex,  
\beeq
\tilde{T}^{ab}_{\mu, \tau}(p,q) = \delta_{ab} (\not p \delta_{\mu \tau} -p_\mu \gamma_\tau) \, .
\eneq

The tree-level expression for the amputated matrix element $\big< {\cal O} \nabla_\mu S_\mu(x) \big>$ 
(using the notation in Eq.~(\ref{a32}) is
\beeq
\bigg(\nabla_\mu S_\mu \bigg)^{(0)}_{amp} \stackrel{FT}{\Longrightarrow}  2 (p - q)_\mu \sigma_{\rho \nu} \gamma_\mu p_\rho \, ,
\label{a46}
\eneq
while the tree-level expression for the amputated matrix element 
$\big< {\cal O} \nabla_\mu T_\mu(x) \big>$ is 
\beeq
\bigg(\nabla_\mu T_\mu \bigg)^{(0)}_{amp} \stackrel{FT}{\Longrightarrow} i (\not p p_\nu - p^2 \gamma_\nu -
\not p q_\nu + p \cdot q \gamma_\nu ) \, .
\label{a47}
\eneq
In our convention, $\nabla_\mu = i (p - q)_\mu $, is the momentum transfer of the operator insertion.

From Eqs.~(\ref{a46},\ref{a47}) it is easy to see that, 
for $p = q $, a condition which would greatly simplify the calculation because implies that the operator insertion 
is at zero momentum, $\bigg(\nabla_\mu S_\mu(x) \bigg)^{(0)}_{amp} = \bigg(\nabla_\mu T_\mu(x) \bigg)^{(0)}_{amp} = 0$. 
So the tree-level of $\nabla_\mu S_\mu$ and $\nabla_\mu T_\mu$ can not be distinguished at zero momentum transfer. 
In order to determine $Z_T$, different tree-level values of $S_\mu$ and $T_\mu$ are needed. 
To differenciate these tree-level values, general external momenta, $p$ and $q$ are required.

The value of the projections over $\gamma_\alpha$ and 
$\gamma_\alpha \gamma_5$ for the different matrix elements in Eq.~(\ref{a8}) has been performed.
Denoting $\frac{1}{4} \mbox{tr}(\gamma_\alpha (\nabla_\mu S_\mu)_{amp})$ the projection
over $\gamma_\alpha$ and 
$\frac{1}{4} \mbox{tr}(\gamma_\alpha \gamma_5 (\nabla_\mu S_\mu)_{amp})$ the projection over 
$\gamma_\alpha \gamma_5$ ($\mbox{tr}$ is the trace over the gamma matrices which should not be confused with $\mbox{Tr}$, the
trace over the color indices), it is easy to demonstrate that 
\beeqa
&& \hspace{-1cm} \frac{1}{4} \mbox{tr} \bigg( \gamma_\alpha \bigg(\nabla_\mu S_\mu \bigg)^{(0)}_{amp} \bigg) \stackrel{FT}{\Longrightarrow}  \nonumber \\
&& 2 i (p_\alpha p_\nu - p_\alpha q_\nu - p^2 \delta_{\alpha \nu} + p \cdot q \delta_{\alpha \nu})
\eneqa
where $\frac{1}{4} \mbox{tr} (\gamma_\mu \gamma_\rho) = \delta_{\mu \rho}$ and 
$\frac{1}{4} \mbox{tr}(\gamma_5 \gamma_\mu  \gamma_\nu \gamma_\rho \gamma_\sigma) = \varepsilon_{\mu \nu \rho \sigma}$, while
\beeq
\frac{1}{4} \mbox{tr} \bigg( \gamma_\alpha \gamma_5 \bigg(\nabla_\mu S_\mu \bigg)^{(0)}_{amp} \bigg) \stackrel{FT}{\Longrightarrow} 
2 i p_\rho q_\sigma \varepsilon_{\nu \alpha \rho \sigma} \, .
\eneq
Also 
\beeqa
&& \hspace{-1cm} \frac{1}{4} \mbox{tr} \bigg( \gamma_\alpha \bigg(\nabla_\mu T_\mu \bigg)^{(0)}_{amp} \bigg) \stackrel{FT}{\Longrightarrow} \nonumber \\
&& i (p_\alpha p_\nu - p_\alpha q_\nu - p^2 \delta_{\alpha \nu} + p \cdot q \delta_{\alpha \nu})
\eneqa
and 
\beeq
\frac{1}{4} \mbox{tr} \bigg( \gamma_\alpha \gamma_5 \bigg(\nabla_\mu T_\mu \bigg)^{(0)}_{amp} \bigg) \stackrel{FT}{\Longrightarrow}  0 \, .
\eneq

Concerning the gauge fixing term, the tree-level value can be read from Eq.~(\ref{a34})
\beeq
- \bigg( \frac{\delta S_{GF}}{\delta \xi(x)}|_{\xi=0} \bigg)^{(0)}_{amp} \stackrel{FT}{\Longrightarrow} -2 i \not p p_\nu 
\label{a57}
\eneq
and the projections are 
\beeq
- \frac{1}{4} \mbox{tr} \bigg( \gamma_\alpha \bigg(\frac{\delta S_{GF}}{\delta \xi(x)}|_{\xi=0} \bigg)^{(0)}_{amp} \bigg) \stackrel{FT}{\Longrightarrow}
-2 i p_\alpha p_\nu
\eneq
and 
\beeq
- \frac{1}{4} \mbox{tr} \bigg( \gamma_\alpha \gamma_5 \bigg(\frac{\delta S_{GF}}{\delta \xi(x)}|_{\xi=0} \bigg)^{(0)}_{amp} \bigg) 
\stackrel{FT}{\Longrightarrow}  0 \, .
\eneq
For the contact terms, the tree-level can be seen directly from Eq.~(\ref{a5}) (with $a \to 0$)
\beeqa
\bigg< \frac{\delta {\cal O} }{\delta \xi(x)}|_{\xi=0} \bigg>^{(0)} &=& 
2 i \delta(x - y) \gamma_\nu \big<  \lambda^a(y) \bar \lambda^b(z) \big> + \nonumber \\
&& \hspace{-1 cm} \delta(x - y)  \big< A_\nu^a(y) \sigma_{\rho \tau} {\cal G}_{\rho \tau}^b(z) \big> 
\label{a35}
\eneqa
or in FT (in the limit $m_0 \to 0$)
\beeq
2 i \gamma_\nu \bigg( \frac{1}{i \not q } \bigg) \delta_{a b} -
2 i p_\rho \frac{1}{p^2} \sigma_{\rho \nu} \delta_{a b} \, .
\eneq
The projections are
\beeqa
&& \hspace{-2cm} \frac{1}{4} \mbox{tr} \bigg( \gamma_\alpha \bigg(\frac{\delta {\cal O}}{\delta \xi(x)}|_{\xi=0} \bigg)^{(0)}_{amp} \bigg) \stackrel{FT}{\Longrightarrow} \nonumber \\
&& 2 i ( p_\alpha q_\nu - p \cdot q \delta_{\nu \alpha} + p^2 \delta_{\alpha \nu}) 
\eneqa
and 
\beeq
\frac{1}{4} \mbox{tr} \bigg( \gamma_\alpha \gamma_5 \bigg(\frac{\delta {\cal O}}{\delta \xi(x)}|_{\xi=0} \bigg)^{(0)}_{amp} \bigg) \stackrel{FT}{\Longrightarrow} 
- 2 i p_\rho q_\sigma \varepsilon_{\nu \alpha \rho \sigma} \, .
\eneq
Finally, the tree-level vertex for the operator in Eq.~(\ref{a58}) is
\beeq
( \tilde{B}_0)^{ab}_\tau(p,q)= i \delta_{ab} (p_\rho p_\tau - p^2 \delta_{\rho \tau}) \gamma_\rho 
\eneq
while the projection is
\beeq
\frac{1}{4} \mbox{tr} \bigg( \gamma_\alpha \big( B_0 \big)^{(0)}_{amp} \bigg) \stackrel{FT}{\Longrightarrow} 
i (p_\alpha p_\nu - p^2 \delta_{\alpha \nu}) \, .
\label{a60}
\eneq
For the operators in Eq.~(\ref{a59}) we have
\beeqa
&& \frac{1}{4} \mbox{tr} ( \gamma_\alpha (B_1)_{amp} ) \stackrel{FT}{\Longrightarrow} i p_\nu q_\alpha \, , \nonumber \\
&& \frac{1}{4} \mbox{tr} ( \gamma_\alpha (B_2)_{amp} ) \stackrel{FT}{\Longrightarrow} i q_\nu q_\alpha \, , \nonumber \\
&& \frac{1}{4} \mbox{tr} ( \gamma_\alpha (B_3)_{amp} ) \stackrel{FT}{\Longrightarrow} i q^2 \delta_{\alpha \nu} 
\label{a62}
\eneqa
and
\beeq
\frac{1}{4} \mbox{tr} \bigg( \gamma_\alpha \gamma_5 \big( B_{0,1,2,3} \big)^{(0)}_{amp} \bigg) \stackrel{FT}{\Longrightarrow} 0 \, .
\label{a61}
\eneq

The renormalization constants can be written as a power of $g_0$
\beeq
Z_{operator} = Z_{operator}^{(0)} + g_0^2 Z_{operator}^{(2)} + \cdots \, ,
\label{a23}
\eneq
and also for the operators, a similar expansion can be done it
\beeq
\big< Operator \big> = \big< Operator \big>^{(0)} + g_0^2 \big< Operator \big>^{(2)} + \cdots \, ,
\label{a24}
\eneq
where $\big< Operator \big>^{(2)}$, is the 1-loop correction
while $\big< Operator \big>^{(0)}$, is the tree-level value.

Substituting Eq.~(\ref{a23}) and Eq.~(\ref{a24}) into Eq.~(\ref{a8}), to order $g_0^2$, we obtain
\beeqa
&& (1 + g_0^2 Z_S^{(2)}) \bigg( \big< {\cal O} \nabla_\mu S_\mu(x) \big>^{(0)} + \nonumber \\
&& \hspace{3 cm} g_0^2 \big< {\cal O} \nabla_\mu S_\mu(x)\big>^{(2)} \bigg) + \nonumber \\
&& g_0^2 Z_T^{(2)} \big< {\cal O} \nabla_\mu T_\mu(x) \big>^{(0)} + \nonumber \\
&& (1 + g_0^2 Z_{CT}^{(2)}) \bigg( \bigg< \frac{ \delta {\cal O}}{\delta \bar \xi(x)}|_{\xi = 0} \bigg>^{(0)} + \nonumber \\ 
&& \hspace{3 cm} g_0^2 \bigg< \frac{ \delta {\cal O}}{\delta \bar \xi(x)}|_{\xi = 0} \bigg>^{(2)} \bigg) - \nonumber \\
&& (1 + g_0^2 Z_{GF}^{(2)}) \bigg( \bigg< {\cal O}  \, \frac{\delta S_{GF}}{\delta \bar \xi(x)}|_{\xi = 0} \bigg>^{(0)} + \nonumber \\ 
&& \hspace{3 cm} g_0^2 \bigg< {\cal O}  \, \frac{\delta S_{GF}}{\delta \bar \xi(x)}|_{\xi = 0} \bigg>^{(2)} \bigg) - \nonumber \\
&& g_0^2 Z_{FP}^{(2)} \bigg< {\cal O} \, \frac{\delta S_{FP}}{\delta \bar \xi(x)}|_{\xi = 0} \bigg>^{(0)} + \nonumber \\
&& g_0^2 \sum_j Z_{B_j}^{(2)} \big< {\cal O} B_j(x) \big>^{(0)} = 0
\label{a25}
\eneqa

At tree-level we have, $Z_S^{(0)}= 1, Z_T^{(0)}= 0, Z_{CT}^{(0)}= 1, 
Z_{GF}^{(0)}= 1, Z_{FP}^{(0)}= 0, Z_{B_i}^{(0)}= 0$, so the lattice WTi is
\beeqa
&& \big< {\cal O} \nabla_\mu S_\mu(x) \big>^{(0)} + 
\bigg< \frac{ \delta {\cal O}}{\delta \bar \xi(x)}|_{\xi = 0} \bigg>^{(0)} - \nonumber \\
&& \bigg< {\cal O}  \, \frac{\delta S_{GF}}{\delta \bar \xi(x)}|_{\xi = 0} \bigg>^{(0)}  = 0 \, ,
\label{a29}
\eneqa
which holds in our lattice calculation. 
Eq.~(\ref{a29}) was previously determined in the continuum \cite{galla2}.
To order $g_0^2$ the lattice WTi is
\beeqa
&& \big< {\cal O} \nabla_\mu S_\mu(x) \big>^{(2)} + 
 Z_S^{(2)} \big< {\cal O} \nabla_\mu S_\mu(x) \big>^{(0)} + \nonumber \\ 
&& Z_T^{(2)} \big< {\cal O} \nabla_\mu T_\mu(x) \big>^{(0)} + \nonumber \\
&& \bigg< \frac{ \delta {\cal O}}{\delta \bar \xi(x)}|_{\xi = 0} \bigg>^{(2)} + 
Z_{CT}^{(2)} \bigg< \frac{ \delta {\cal O}}{\delta \bar \xi(x)}|_{\xi = 0} \bigg>^{(0)} - \nonumber \\
&& Z_{GF}^{(2)} \bigg< {\cal O}  \, \frac{\delta S_{GF}}{\delta \bar \xi(x)}|_{\xi = 0} \bigg>^{(0)} - 
\bigg< {\cal O}  \, \frac{\delta S_{GF}}{\delta \bar \xi(x)}|_{\xi = 0} \bigg>^{(2)} - \nonumber \\
&&  \sum_j Z_{B_j}^{(2)} \big< {\cal O} B_j(x) \big>^{(0)} = 0 \, .
\label{a26}
\eneqa
Notice that the Faddeev-Popov term 
$\bigg< {\cal O} \, \frac{\delta S_{FP}}{\delta \bar \xi(x)}|_{\xi = 0} \bigg>^{(0)}$ 
in Eq.~(\ref{a25}), is already $O(g_0^2)$ (see Appendix~\ref{appendixVERTICES}) and
does not contribute to 1-loop order. 
In Fig.~\ref{Fig1}, the Feynman diagrams for $\big< {\cal O} \nabla_\mu S_\mu(x) \big>^{(2)}$
and $\bigg< {\cal O}  \, \frac{\delta S_{GF}}{\delta \bar \xi(x)}|_{\xi = 0} \bigg>^{(2)}$ are shown, while 
in Fig.~\ref{Fig2}, the non-zero contribution to contact terms are presented.

Let us substitute the tree-level values of the operators in Eq.~(\ref{a26}) using the projections 
over $\gamma_\alpha$,
\beeqa
&& \frac{1}{4} \mbox{tr} \bigg( \gamma_\alpha \bigg(\nabla_\mu S_\mu \bigg)^{(2)}_{amp} \bigg) + 
\nonumber \\
&& Z_S^{(2)} 2 i (p_\alpha p_\nu- p_\alpha q_\nu - p^2 \delta_{\alpha \nu} + p \cdot q \delta_{\alpha \nu}) + \nonumber \\
&& Z_T^{(2)} i (p_\alpha p_\nu - p_\alpha q_\nu - p^2 \delta_{\alpha \nu} + p \cdot q \delta_{\alpha \nu}) + \nonumber \\ 
&& \frac{1}{4} \mbox{tr} \bigg(\gamma_\alpha \bigg(\frac{\delta {\cal O}}{\delta \bar \xi(x)}|_{\xi = 0} \bigg)_{amp}^{(2)} \bigg) + \nonumber \\
&& Z_{CT}^{(2)} 2 i (p_\alpha q_\nu - p \cdot q \delta_{\alpha \nu} + p^2 \delta_{\alpha \nu} ) - \nonumber \\
&& Z_{GF}^{(2)} 2 i p_\alpha p_\nu - \nonumber \\ 
&& \frac{1}{4} \mbox{tr} \bigg( \gamma_\alpha \bigg(\frac{\delta S_{GF}}{\delta \bar \xi(x)}|_{\xi = 0} \bigg)_{amp}^{(2)} \bigg) + \nonumber \\
&& \frac{1}{4} Z_{B_j}^{(2)} \mbox{tr} \big< \gamma_\alpha {\cal O} B_j \big>^{(0)} = 0 \, ,
\label{a27}
\eneqa
and the projections over $\gamma_\alpha \gamma_5$,
\beeqa
&& \frac{1}{4} \mbox{tr} \bigg( \gamma_\alpha \gamma_5 \bigg( \nabla_\mu S_\mu \bigg)_{amp}^{(2)} \bigg) + \nonumber \\ 
&& Z_S^{(2)} 2 i p_\rho q_\sigma \varepsilon_{\nu \alpha \rho \sigma} - \nonumber \\ 
&& Z_{CT}^{(2)} 2 i p_\rho q_\sigma \varepsilon_{\nu \alpha \rho \sigma} + \nonumber \\
&& \frac{1}{4} \mbox{tr} \bigg( \gamma_\alpha \gamma_5 \bigg(\frac{ \delta {\cal O}}{\delta \bar \xi(x)}|_{\xi = 0} \bigg)_{amp}^{(2)} \bigg) - \nonumber \\
&& \frac{1}{4} \mbox{tr} \bigg( \gamma_\alpha \gamma_5 \bigg( \frac{\delta S_{GF}}{\delta \bar \xi(x)}|_{\xi = 0} \bigg)_{amp}^{(2)} \bigg) + \nonumber \\ 
&& \frac{1}{4} Z_{B_j}^{(2)} \mbox{tr} \big< \gamma_\alpha \gamma_5 {\cal O} B_j \big>^{(0)} = 0 \, .
\label{a28}
\eneqa
Our claim is that, in order to calculate $Z_T^{(2)}$ we can substitute 
$\frac{1}{4} Z_{B_i}^{(2)} \mbox{tr} \big< \gamma_\alpha {\cal O} B_i \big>^{(0)} \to 
Z_{B_0}^{(2)} i (p_\alpha p_\nu - p^2 \delta_{\alpha \nu}) + Z_{B_1}^{(2)} i p_\nu q_\alpha + 
Z_{B_2}^{(2)} i q_\nu q_\alpha + Z_{B_3}^{(2)} i q^2 \delta_{\alpha \nu} $, and 
$\frac{1}{4} Z_{B_i}^{(2)} \mbox{Tr} \big< \gamma_\alpha \gamma_5 {\cal O} B_j \big>^{(0)} \to 0$,
where $Z_{B_j}$ correspond to the renormalization constant in Eq.~(\ref{a60},\ref{a62},\ref{a61}).
No other $Z_{B_j}$ are needed, because there are no other
$B_j$'s that would contribute with the same Lorentz structures appearing in the tree-level of 
Eqs.~(\ref{a27},\ref{a28}). 
 
Each matrix element in Eqs.~(\ref{a27},\ref{a28}) has been calculated for general 
$p$ and $q$ (off-shell regime).
To deal with the IR divergencies and rinormalize to 1-loop order the Kawai procedure is used
\cite{kawai}, with the help of tabulated results in \cite{martinelli,haris}.
Once $p$ and $q$ have been extracted from the propagators through the Kawai procedure, the rest
of the integral depend on the loop momenta which is numerically integrated. A similar renormalization procedure  
has been used to calculate the 3-loop beta function in QCD with Wilson fermions \cite{feo6} and the 3-loop
free-energy in QCD with Wilson fermions \cite{feo7,feo8} (for a complete study of the off-shell WTi 
in QCD see \cite{menotti}). 
Tipically, each matrix element contains $\approx 1000 $ terms 
(in particular dilogarithms functions depending on both external momenta which coming from the diagrams with 
three propagators in Fig.~\ref{Fig1}).
After the numerical integration, one can simplify the results in order to read the value of 
$Z_T$ by setting 
\beeq
p^2 = q^2  \, \, \, \, \mbox{and} \, \, \, \, p \cdot q = 0 \, ,
\label{a51}
\eneq
(see Appendix~\ref{appendixOFFSHELL}).
This is still an off-shell condition (because even if $p^2 = q^2$, there are no other condition on this 
expression, i.e., $q^2 = 0$), but drastically reduces the number and difficulty of the expressions 
(for example, the dilogarithm terms simplify).
 
Let us introduce, for simplicity, the notation,
$\Delta \equiv {\cal O} \nabla_\mu S_\mu(x) + \frac{ \delta {\cal O}}{\delta \bar \xi(x)}|_{\xi = 0} -
{\cal O}  \, \frac{\delta S_{GF}}{\delta \bar \xi(x)}|_{\xi = 0} $.
Using Eq.~(\ref{a51}), we get the following dependence on $p$ and $q$ for
$\mbox{tr} \big< \gamma_\alpha \Delta \big>^{(2)} $ 
and $\mbox{tr} \big< \gamma_\alpha \gamma_5 \Delta \big>^{(2)} $,
\beeqa
\mbox{tr} \big< \gamma_\alpha \Delta \big>^{(2)} &\stackrel{FT}{\Longrightarrow} & A_1 q^2 \hat p_\alpha \hat p_\nu + 
A_2 q^2 \hat p_\alpha \hat q_\nu + \nonumber \\ 
&& \hspace{-2cm} (A_3 + M_3) q^2  \delta_{\alpha \nu} + M_1 q^2 \hat p_\nu \hat q_\alpha +
 M_2 q^2 \hat q_\alpha \hat q_\nu + \nonumber \\ 
&&  \hspace{-2cm} P_1 q^2 \hat p_\nu^2 \delta_{\nu \alpha} + 
P_2 q^2 \hat q_\nu^2 \delta_{\nu \alpha} + \cdots
\label{a36}
\eneqa
and 
\beeq
\mbox{tr} \big< \gamma_\alpha \gamma_5 \Delta \big>^{(2)} \stackrel{FT}{\Longrightarrow}  
A_4 q^2 \hat p_\rho \hat q_\sigma \varepsilon_{\nu \alpha \rho \sigma} \, ,
\label{a37}
\eneq
where the dots in Eq.~(\ref{a36}) indicate that because the semplification in Eq.~(\ref{a51}) is used, some 
momenta dependence are missing or mixed with others, i.e., $ p \cdot q \delta_{\nu \alpha} $ does 
not appear, while $p^2 \delta_{\nu \alpha}$, is mixed with $q^2 \delta_{\nu \alpha}$, 
(see Appendix~\ref{appendixOFFSHELL} for notation). 

It is also interesting to see the Lorentz structure of the supercurrent,
\beeqa
\mbox{tr} \big< \gamma_\alpha S_\mu \big>^{(2)} &\stackrel{FT}{\Longrightarrow} & 
N_1 q \, \hat p_\mu \hat p_\nu \hat p_\alpha +  N_2 q \, \hat q_\mu \hat p_\nu \hat p_\alpha + \nonumber \\  
&& \hspace{-2cm} N_3 q \, \hat p_\mu \hat q_\nu \hat p_\alpha +  N_4 q \, \hat q_\mu \hat q_\nu \hat p_\alpha +
N_5 q \, \hat p_\mu \hat p_\nu \hat q_\alpha  + \nonumber \\
&& \hspace{-2cm} N_6 q \, \hat q_\mu \hat p_\nu \hat q_\alpha + N_7 q \, \hat p_\mu \hat q_\nu \hat q_\alpha +  
N_8 q \, \hat q_\mu \hat q_\nu \hat q_\alpha + \nonumber \\
&& \hspace{-2cm} Q_1 q \, \hat p_\alpha \delta_{\mu \nu} + Q_2 q \, \hat q_\alpha \delta_{\mu \nu} + 
Q_3 q \, \hat p_\nu \delta_{\mu \alpha} + \nonumber \\ 
&& \hspace{-2cm}  Q_4 q \, \hat q_\nu \delta_{\mu \alpha} + Q_5 q \, \hat p_\mu \delta_{\nu \alpha} + 
Q_6 q \, \hat q_\mu \delta_{\nu \alpha} + \nonumber \\
&& \hspace{-2cm} R_1 q \, \hat p_\mu \delta_{\mu \nu \alpha} + R_2 q \, \hat q_\mu \delta_{\mu \nu \alpha} + \cdots
\label{a63}
\eneqa
where the coefficient $A_i,M_j,Q_k$, are tipically of the form
\beeq
(C_n + C_m \mbox{Ln}(a^2 q^2) )
\eneq
while $P_i,N_j,R_k$, do not contain $ \mbox{Ln}(a^2 q^2)$ terms. Here,
$C_n$ are lattice constants or numbers coming from the numerical integration and $C_m$ are rational
numbers coming from the Kawai procedure.
Notice that the Lorentz structures multiplying $P_i,R_k$ in Eq.~(\ref{a36},\ref{a63}) are non-Lorentz covariant,
even in the continuum limit $(a \to 0)$. 

From Eqs.~(\ref{a27},\ref{a28},\ref{a36},\ref{a37}) the following conditions can be derived
\beeqa
&& \hspace{-0.4cm} A_1 = -2 i Z_S^{(2)} - i Z_T^{(2)} + 2 i Z_{GF}^{(2)} -i Z_{B_0}^{(2)} \, , \nonumber \\
&& \hspace{-0.4cm} A_3 + M_3 = 2 i Z_S^{(2)} + i Z_T^{(2)} - 2 i Z_{CT}^{(2)} + i Z_{B_0}^{(2)} -i Z_{B_3}^{(2)} \, ,\nonumber  \\
&& \hspace{-0.4cm} M_1 = -i Z_{B_1}^{(2)} \, , \nonumber \\
&& \hspace{-0.4cm} M_2 = -i Z_{B_2}^{(2)} \, 
\eneqa
and 
\beeqa
&& \hspace{-0.4cm} A_2 = 2 i Z_S^{(2)} + i Z_T^{(2)} - 2 i Z_{CT}^{(2)} \, , \nonumber \\
&& \hspace{-0.4cm} A_4 = -2 i Z_S^{(2)} + 2 i Z_{CT}^{(2)} \, .
\label{a56}
\eneqa
The last two conditions can be explicitly solved for $Z_T^{(2)}$,
\beeq
Z_T^{(2)} = -i A_2 - i A_4 \, .
\label{a55}
\eneq
Eq.~(\ref{a55}) is the only possible solution of the system (\ref{a27},\ref{a28}) 
for $Z_T^{(2)}$. Our result is $Z_T^{(2)}|_{1-loop} = 0.664$. 
A VEGAS Monte Carlo routine to perform the 1 loop integration 
with 200 millions of points, using the GNU Scientific Library 
\footnote{http//:www.gnu.org/software/gsl/} is used.
To estimate the error we take the value given by the program which is $\approx 10^{-5}$ for each integral. 
The calculation, once $p$ and $q$ has been extracted from the propagators,
involves around 1300 different 1-loop integrals. For each diagram tipically we have 100 different integrals.
That means that the error is around $10^{-3}$. 

Let us compare our perturbative result with the numerical one \cite{farchioni}, $ Z_T^{NUM} \equiv Z_T/Z_S = 0.185(7)$.
One has to observe that the definition used here for $S_\mu$ is not the same as in \cite{farchioni}.
It is easy to demonstrate that $Z_T^{NUM} = \frac{1}{2} Z_T^{PT}$ \cite{farchioni2}. 
To compare with the numerical results one as to divide the perturbative value
by 2 which gives, $Z_T^{PT} = \frac{1}{2} Z_T^{(2)}|_{1-loop} = 0.332$.
We are currently increasing the precision of the numerical integration to 400 milion of points.
A detailed presentation of the results in Eqs.~(\ref{a36},\ref{a37}) together with the result of 
each diagram is under way \cite{feo3}. 

\section{Discussion and conclusions}
\label{sec5}
\noindent
In this paper, the SUSY WTi in 1-loop LPT has been investigated. 
A general procedure in order to get the renormalization constant for the supercurrent has been 
presented.
In LPT it is possible to determine the value of the renormalization 
constant for the supercurrent from the off-shell regime of the SUSY WTi.
The computation of each matrix elements of the WTi has been carried out using the symbolic language 
Mathematica. The programs were completely wroted by the author together with the numerical code 
used for the integration.
All the contributions have been calculated in the off-shell regime, and in order to get the 
value of the renormalization constant, a simplification in the external momenta (which still 
keeps the off-shell regime) has been applyed.
We are currently increasing the precision of the numerical integration and a detailed presentation of the results
is the subject of a forthcoming paper \cite{feo3}.
A reasonable good agreement of our perturbative result for the renormalization constant, $Z_T^{PT} = 0.332$,
in comparison with the numerical one, $Z_T^{NUM} = 0.185(7)$, has been achieved, taking in 
consideration the fact that in the numerical simulation $g_0^2 = 4/2.3$, which still corresponds to the
non-perturbative region. 
We observe that, at least at 1-loop order in perturbation theory, $Z_T$ is finite.
This result may have some theoretical implications which we are currently investigating. Also, the 
determination of $Z_S$, using another kind of gamma projections is under investigation.
It would be interesting to calculate $Z_S$ in order to check the trace anomaly and the exact renormalization 
expression for Eq.~(\ref{a50}).
An important point to stress here is that, even in the continuum limit, we observe in Eq.~(\ref{a36})
Lorentz breaking terms which coming from the fact that we substituted $X_S$ by the Eq.~(\ref{a45}).
It would be interesting to see whether Eq.~(\ref{a36}) is the continuum off-shell WTi.
The nice point is that, once the $Z_T$ has been determined, we can impose the 
on-shell condition on the gluino mass. The Lorentz breaking terms cancel out from Eq.~(\ref{a36})
and the continuum WTi is recovered.
At least to 1-loop order, we do not observe a SUSY anomaly in $N=1$ SYM, altough 
a more carefully study is required.
  
\acknowledgments
It is a pleasure to thank Marisa~Bonini, Massimo~Campostrini, Matteo~Beccaria, Giuseppe~Burgio and 
Roberto De Pietri for useful and stimulating discussions.
A.~F. is indebt with Federico~Farchioni, Tobias~Galla, Claus~Gebert, Robert~Kirchner, Istv\'an~Montvay,
Gernot~M\"unster, Roland~Peetz and Anastassios~Vladikas, for collaboration in earlier works, 
from which their contributions to this paper benefit.
This work was partially funded by the Enterprise-Ireland grant SC/2001/307.

\appendix
\section{Perturbative Calculation} 
\label{appendixPERT}
\noindent
In this appendix we follow the lines of \cite{feo,galla2}. 
The lattice SUSY transformations of the gauge field $A_\mu(x)$ 
are not equal to the continuum ones. On the lattice the transformation 
of the gauge link $U_\mu(x)$ determines the transformation
properties of $A_\mu(x)$. Writing the link variable as
\beeq
U_\mu(x) = e^{-a A_\mu \left(x + \frac{a}{2} \hat \mu \right)} \, ,
\eneq
for the SUSY transformations of the gauge link we use the symmetric choice
\cite{galla} 
\beeqa
\delta U_\mu(x) &=& - a g_0 U_\mu(x) \bar \xi(x) \gamma_\mu \lambda(x) \nonumber \\
&& - a g_0 \bar \xi(x + a \hat \mu) \gamma_\mu \lambda(x + a \hat \mu) U_\mu(x) \, . \nonumber 
\eneqa
These two equations determine the transformation behavior of the field $A_\mu(x)$ \cite{feo}.
The FT for the gauge field is defined as the usual way
\beeq
A^b_\mu(x) = \int d^4k \tilde{A}^b_\mu(k) e^{ik \cdot (x + \frac{a}{2} \hat \mu) } \, .
\eneq
Collecting all terms until order $g_0^2$ we can write down the variation of the 
gauge field $A_\mu^b(x)$ as \cite{feo}
\beeqa
\delta A_\mu^b(x) & = &  i \bigg(\bar \xi(x) \gamma_\mu \lambda^b(x) +
\bar \xi(x + a \hat\mu) \gamma_\mu \lambda^b(x + a \hat\mu) \bigg) + \nonumber \\
&& \hspace{-1.6 cm} \frac{i}{2} a g_0 f_{abc} 
\bigg( \bar \xi(x) \gamma_\mu  \lambda^c(x) - \bar \xi(x + a \hat\mu) 
\gamma_\mu \lambda^c(x + a \hat\mu) \bigg) A_\mu^a - \nonumber \\
&& \hspace{-1.4 cm} \frac{i}{24} a^2 g_0^2 \bigg( 2 \delta_{ab} \delta_{cd} 
- \delta_{ac} \delta_{bd} - \delta_{ad} \delta_{bc} \bigg) 
A_\mu^c A_\mu^d \times  \nonumber \\
 && \hspace{-1.4 cm} \times \bigg( \bar \xi(x) \gamma_\mu \lambda^a(x) + \bar \xi(x + a \hat\mu) 
\gamma_\mu \lambda^a(x + a \hat\mu) \bigg)
\label{a9}
\eneqa
which reduces to the continuum SUSY transformation 
$ \delta A_\mu^a(x) = 2 i \bar \xi \gamma_\mu \lambda^a(x) $ 
in the continuum limit $a \rightarrow 0$. Because in this paper we fix $N_c = 2$, 
some semplifications appear
\beeqa
\mbox{Tr} \bigg\{ T^a T^b T^c T^d \bigg\} & = & \frac{1}{8} (\delta_{ab} \delta_{cd} - 
f_{abe} f_{cde}) \, , \nonumber \\
f_{abe} f_{cde}  & = &  (\delta_{ac} \delta_{bd} - \delta_{ad} \delta_{bc})
\, .
\eneqa

Using the Eq.~(\ref{a9}), it is possible to determine the differents pieces of the WTi in Eq.~(\ref{a8}),
i.e., the contact terms, the gauge fixing term and the Faddeev-Popov term.
They are necessary in order to calculate the Feynman rules for 1-loop order calculation.
In appendix~\ref{appendixVERTICES}, the vertices coming from these pieces are presented together with 
the ones coming from the supercurrent.

\section{Vertices}
\label{appendixVERTICES}
\noindent
Let us determine the contact terms, $\frac{\delta {\cal O}} {\delta \bar \xi(x)}|_{\xi = 0}$. 
First of all, the variation of the operator insertion, ${\cal O} = A_\nu^a(y) \bar \lambda^b(z)$, is
\beeqa
\delta {\cal O} = \delta A_\nu^a(y) \bar \lambda^b(z) + 
A_\nu^a(y) \delta \bar \lambda^b(z) \, .
\label{a10}
\eneqa
Substituting (\ref{a9}) into (\ref{a10}), after some algebra, we obtain 
\beeqa
&&  \bigg< \frac{\delta {\cal O}}{\delta \bar \xi(x)}|_{\xi=0}\bigg> = \nonumber \\
&& = i \delta(x - y) \gamma_\nu \bigg<  \lambda^a(y) \bar \lambda^b(z) \bigg> + \nonumber \\ 
&& i \delta(x - y - a \hat \nu)  \gamma_\nu \bigg<  \lambda^a(y + a \hat \nu) 
\bar \lambda^b(z) \bigg> + \nonumber \\
&& \frac{i}{2} a g_0 f_{dac} \delta(x - y) 
\gamma_\nu \bigg< \lambda^c(y) A_\nu^d(y) \bar \lambda^b(z) \bigg> - \nonumber \\
&& \frac{i}{2} a g_0 f_{dac} \delta(x - y - a \hat \nu ) 
 \gamma_\nu \bigg< \lambda^c(y + a \hat \nu) A_\nu^d(y) \bar \lambda^b(z) \bigg> - \nonumber \\
&& \frac{i}{24} a^2 g_0^2 \bigg( 2 \delta_{a e} \delta_{c d} - \delta_{e c} \delta_{a d}  
- \delta_{e d} \delta_{a c} \bigg) \delta(x - y) \times \nonumber \\ 
&& \hspace{0.9 cm} \times \bigg< A_\nu^c(y) A_\nu^d(y) \lambda^e(y) \bar \lambda^b(z) \bigg> \nonumber \\
&&  - \frac{i}{24} a^2 g_0^2 \bigg( 2 \delta_{a e} \delta_{c d} - \delta_{e c} \delta_{a d}  
- \delta_{e d} \delta_{a c} \bigg) \delta(x - y - a \hat \nu ) \times \nonumber \\ 
&& \hspace{0.9 cm} \times \bigg< A_\nu^c(y) A_\nu^d(y) \lambda^e(y + a \hat \nu) \bar \lambda^b(z) \bigg> + \nonumber \\
&& \delta(x - y)  \bigg< A_\nu^a(y) \sigma_{\rho \tau} {\cal G}_{\rho \tau}^b(z) \bigg> \, ,
\label{a5}
\eneqa
where ${\cal G}_{\rho \tau}(z) = -i g_0 {\cal G}_{\rho \tau}^b(z) T^b$.  

\noindent
The part of the lattice action corresponding to the gauge fixing is defined as
\beeqa
\label{a16}
S_{GF} &=& \frac{a^2}{2} \sum_x \bigg( \sum_\rho ( A_\rho^c(x) - 
A_\rho^c(x - a \hat \rho) ) \bigg)^2 \nonumber \\
&& = \frac{a^4}{2} \sum_x \bigg( \sum_\rho \nabla_\rho^{back} A_\rho^c(x) \bigg)^2  \\
&& = \frac{a^4}{2} \sum_x \bigg( \sum_\rho \nabla_\rho^{back} A_\rho^c(x) \bigg) 
                 \bigg( \sum_\tau \nabla_\tau^{back} A_\tau^c(x) \bigg) \nonumber 
\eneqa 
where $\nabla_\rho^{back} f(x) = \frac{1}{a} \, \bigg(f(x) - f(x - a \hat \rho) \bigg)$.
\noindent
The variation of the gauge fixing term (\ref{a16}) can be written as
\beeq
\delta S_{GF} = a^4 \sum_{x} \bigg( \sum_\rho \nabla_\rho^{back} \delta A_\rho^c(x) \bigg)
\bigg( \sum_\tau \nabla_\tau^{back} A_\tau^c(x) \bigg) \, . 
\eneq
\noindent
This results in the contribution of the gauge fixing term into the WTi as
\beeqa
&& - \bigg< {\cal O} \, \frac{\delta S_{GF}}{\delta \xi(x)}|_{\xi=0} 
\bigg> = \nonumber \\
&& i \gamma_\rho \bigg[ \bigg< \lambda^c(x) \nabla_\rho^{forw} \nabla_\tau^{back}
\bigg( A_\tau^c(x) + \nonumber \\ 
&& \hspace{2 cm} A_\tau^c(x - a \hat \rho )\bigg) A_\nu^a(y) \bar \lambda^b(z) 
\bigg> + \nonumber \\  
&& \frac{i}{2} a g_0 f_{ecf} \gamma_\rho \bigg< \lambda^f(x) \bigg( 
A_\rho^e \nabla_\rho^{forw} \nabla_\tau^{back} A_\tau^c(x) - \nonumber \\ 
&& A_\rho^e(x - a \hat \rho) \nabla_\rho^{forw}
\nabla_\tau^{back} A_\tau^c(x - a \hat \rho) \bigg) A_\nu^a(y) \bar 
\lambda^b(z) \bigg> - \nonumber \\
&& \frac{i}{24} a^2 g_0^2 \gamma_\rho \bigg( 2 \delta_{ec} \delta_{fd} - 
\delta_{ef} \delta_{cd} - \delta_{ed} \delta_{fc} \bigg) \times \nonumber \\ 
&& \hspace{2 cm} \times \bigg< \lambda^e(x)
\bigg( A_\rho^f A_\rho^d \nabla_\rho^{forw} \nabla_\tau^{back} A_\tau^c(x) 
 + \nonumber \\
&& A_\rho^f(x - a \hat \rho) A_\rho^d(x - a \hat \rho)
\nabla_\rho^{forw} \nabla_\tau^{back} A_\tau^c(x - a \hat \rho) \bigg) \times \nonumber \\
&& \hspace{2 cm} \times A_\nu^a(y) \bar \lambda^b(z) \bigg> \bigg] \, ,
\label{a6}
\eneqa
\noindent
where $\nabla_\rho^{forw} f(x) = \frac{1}{a} \, \big(f(x - a \hat \rho) - f(x) \big)$.

\noindent
Finally, the expansion of the Faddeev Popov action can be written as 
\beeqa
S_{FP} & = & a^2 \sum_{x,\rho^{<}_{>} 0} \bigg[ \bar \eta^a(x) \bigg(
\delta_{ab} + \frac{1}{2} a g_0 A_\rho^c f^{acb} + \nonumber \\ 
&& \frac{1}{12} a^2 g_0^2 A_\rho^c A_\rho^d f^{ace} f^{edb} \bigg) \eta^b(x) - \nonumber \\  
&& \bar \eta^a(x) \bigg(\delta_{ab} - \frac{1}{2} a g_0 A_\rho^c f^{acb} + \nonumber \\
&& \hspace{-0.8cm} \frac{1}{12} a^2 g_0^2 A_\rho^c A_\rho^d f^{ace} f^{edb} \bigg) 
\eta^b(x + a \hat \rho) \bigg] 
\eneqa
and the contribution of the Faddeev-Popov term into the WTi is 
\beeqa
&& - \bigg< {\cal O} \, 
\frac{\delta S_{FP}}{\delta \bar \xi(x)}|_{\xi=0} \bigg> = \nonumber \\ 
&& - \frac{i g_0}{2 a} f^{gch} \sum_\rho \gamma_\rho \bigg< \nonumber \\
&& \bigg[ \bar \eta^g(x) \lambda^c(x) \bigg( \eta^h(x) + \eta^h(x + a \hat \rho) \bigg) + \nonumber \\
&& \bar \eta^g(x - a \hat \rho) \lambda^c(x) \bigg( \eta^h(x - a \hat \rho) + \eta^h(x) \bigg) - \nonumber \\
&& \bar \eta^g(x + a \hat \rho) \lambda^c(x) \bigg( \eta^h(x + a \hat \rho) + \eta^h(x) \bigg) + \nonumber \\
&& \bar \eta^g(x) \lambda^c(x) \bigg( \eta^h(x) + \eta^h(x - a \hat \rho) \bigg) \bigg]
 A_\nu^a(y) \bar \lambda^b(z) \bigg> - \nonumber \\
&& \frac{i g_0^2}{4} f^{gch} f^{dcf} \sum_\rho \gamma_\rho \bigg< \nonumber \\ 
&& \bigg[ \bar \eta^g(x) \lambda^f(x) A_\rho^d \bigg( \eta^h(x) + \eta^h(x + a \hat \rho) \bigg) - \nonumber \\ 
&& \bar \eta^g(x - a \hat \rho) \lambda^f(x) A_\rho^d(x - a \hat \rho) 
\bigg( \eta^h(x - a \hat \rho) + \eta^h(x) \bigg) - \nonumber \\ 
&& \bar \eta^g(x + a \hat \rho) \lambda^f(x) A_\rho^d \bigg( \eta^h(x + a \hat \rho) + \eta^h(x) \bigg) + \nonumber \\
&& \bar \eta^g(x) \lambda^f(x) A_\rho^d(x - a \hat \rho) 
\bigg( \eta^h(x) + \eta^h(x - a \hat \rho) \bigg) \bigg] \times \nonumber \\ 
&& \hspace{2 cm} \times A_\nu^a(y) \bar \lambda^b(z) \bigg> - \nonumber \\
&& \frac{i g_0^2}{12} f^{gce} f^{edh} \sum_\rho \gamma_\rho \bigg< \nonumber \\ 
&& \bigg[ \bar \eta^g(x) \bigg( \lambda^c(x) A_\rho^d + A_\rho^c \lambda^d(x) \bigg) 
\bigg( \nonumber \\ 
&& \hspace{2 cm} \eta^h(x) - \eta^h(x + a \hat \rho) \bigg) + \nonumber \\
&& \bar \eta^g(x - a \hat \rho) \bigg( \lambda^c(x) A_\rho^d(x - a \hat \rho) + \nonumber \\
&& \hspace{1.6 cm} A_\rho^c(x - a \hat \rho) \lambda^d(x) \bigg) 
\bigg( \eta^h(x - a \hat \rho) - \eta^h(x) \bigg) + \nonumber \\
&& \bar \eta^g(x + a \hat \rho) \bigg( \lambda^c(x) A_\rho^d(x) + 
A_\rho^c(x) \lambda^d(x) \bigg) \bigg( \nonumber \\ 
&& \hspace{2 cm} \eta^h(x + a \hat \rho) - \eta^h(x) \bigg) + \nonumber \\ 
&& \bar \eta^g(x) \bigg( \lambda^c(x) A_\rho^d(x - a \hat \rho) + 
A_\rho^c(x - a \hat \rho) \lambda^d(x) \bigg) \bigg( \nonumber \\ 
&& \hspace{1.4 cm} \eta^h(x) - \eta^h(x - a \hat \rho) \bigg) \bigg] A_\nu^a(y) \bar \lambda^b(z) \bigg> \, ,
\label{a7}
\eneqa
where $A_\rho^c \equiv A_\rho^c(x + \frac{a \hat \rho}{2})$.
\noindent
It is possible to calculate the vertices and the corresponding Feynman diagrams, up to order $g_0^2$, from 
Eq.~(\ref{a5},\ref{a6},\ref{a7}) in FT.

Regarding the contact terms in Eq.~(\ref{a5}), all the contributions to order $g_0^2$ are zero
except for the last line of Eq.~(\ref{a5}). The corresponding non-zero Feynman diagrams
are shown in Fig~\ref{Fig2}. The vertices used here are, the two-gluons vertex,
\beeqa
\tilde{G}_{2 \tau \rho}^{abc}(k_1,k_2) &= & - \frac{1}{2} g_0 f_{abc} \bigg\{ 
\sigma_{\rho \tau } \bigg[ \nonumber \\ 
&& \hspace{-2 cm} \cos(\frac{k_{1 \rho}a}{2} + \frac{k_{2 \rho}a}{2}) \cos(\frac{k_{1 \rho}a}{2}) 
\cos(\frac{k_{1 \tau}a}{2} + k_{2 \tau}a) + \nonumber \\  
&& \hspace{-2 cm} \cos(\frac{k_{1 \tau}a}{2} + \frac{k_{2 \tau}a}{2}) \cos(\frac{k_{2 \tau}a}{2}) 
       \cos(\frac{k_{2 \rho}a}{2} + k_{1 \rho}a) - \nonumber \\ 
&& \hspace{-2 cm} \sin(\frac{k_{1 \rho}a}{2} + \frac{k_{2 \rho}a}{2}) \sin(\frac{k_{1 \rho}a}{2}) 
               \cos(\frac{k_{1 \tau}a}{2}) - \nonumber \\
&& \hspace{-2 cm} \sin(\frac{k_{1 \tau}a}{2} + \frac{k_{2 \tau}a}{2}) \sin(\frac{k_{2 \tau}a}{2}) 
       \cos(\frac{k_{2 \rho}a}{2}) \bigg] - \nonumber \\ 
&& \hspace{-2 cm} \delta_{\rho \tau} \sum_\alpha  \sigma_{\alpha \tau} 
 \sin(\frac{k_{1 \tau}a}{2} + \frac{k_{2 \tau}a}{2}) \bigg(\sin(k_{1 \alpha}a) - \nonumber \\ 
&& \hspace{2 cm} \sin(k_{2 \alpha }a) \bigg) \bigg\} \, , 
\eneqa
and the three-gluons vertex, which we do not reported here and gives a zero contrbution to the last diagram of Fig.~\ref{Fig2}.

For the gauge fixing terms in Eq.~(\ref{a6}), we need the vertex with one-gluon-one-gluino (which is similar to Eq.~(\ref{a57}) in the continuum limit,  
\beeq
\widetilde{GF}_{1 \rho \tau}^{ab}(p,k) = - \frac{4 i}{a^2} \delta_{ab} \gamma_\rho \sin(k_\rho a) \sin(\frac{k_\tau a}{2}) \, ,
\label{a34}
\eneq
the vertex with two-gluons-one-gluino 
\beeqa
\widetilde{GF}_{2 \tau \rho}^{fce}(p,q,k) &=& \nonumber \\ 
&& \hspace{-2 cm} \frac{2 g_0}{a} f_{ecf} \bigg\{ \gamma_\rho \sin(\frac{k_\rho a}{2} + \frac{q_\rho a}{2}) 
\sin(\frac{q_\rho a}{2}) \sin(\frac{q_\tau a}{2}) - \nonumber \\ 
&& \hspace{-2 cm} \gamma_\tau \sin(\frac{k_\tau a}{2} + \frac{q_\tau a}{2}) \sin(\frac{k_\tau a}{2}) \sin(\frac{k_\rho a}{2}) \bigg\} 
\eneqa
and finally the three-gluons-one-gluino vertex (non-symmetrized) 
\beeqa
\widetilde{GF}_{3 \rho \sigma \tau}^{efdc}(p,k,q,t) &=& \nonumber \\ 
&& \hspace{-2.2 cm} - \frac{1}{3} g_0^2 \bigg( 2 \delta_{ec} \delta_{fd} - \delta_{ef} \delta_{cd} - 
\delta_{ed} \delta_{fc} \bigg) \gamma_\rho \delta_{\rho \sigma} \times \nonumber \\
&& \hspace{-2.8 cm} \times \sin(\frac{k_\rho a}{2} + \frac{q_\rho a}{2} + 
\frac{t_\rho a}{2}) \sin(\frac{t_\rho a}{2}) \sin(\frac{t_\tau a}{2})\, .
\eneqa

For the Faddeev-Popov terms in Eq.~(\ref{a7}), we need one-gluino-ghost-antighost vertex
\beeqa
\hspace{-1 cm} \widetilde{FP}_{\rho}^{cgh}(p,-q,k) &=& \nonumber \\ 
&& \hspace{-2.8 cm} \frac{4 g_0}{a} f^{gch} \sum_\rho \gamma_\rho \cos(\frac{k_\rho a}{2}) 
\sin(\frac{q_\rho a}{2}) \cos(\frac{k_\rho a}{2} - \frac{q_\rho a}{2}) 
\label{a38}
\eneqa
and one-gluino-one-gluon-ghost-antighost vertex
\beeqa
\hspace{-1 cm} \widetilde{FP}_{1 \rho}^{cdgh}(p,t,-q,k) & =& \nonumber \\ 
&& \hspace{-2.6 cm} - \frac{2 i}{3} g_0^2 (f^{gce} f^{edh} + f^{gde} f^{ech}) \gamma_\rho \bigg\{ \nonumber \\
&& \hspace{-2.6 cm} \sin(\frac{k_\rho a}{2}) \sin(\frac{q_\rho a}{2}) \cos(\frac{k_\rho a}{2} + \frac{t_\rho a}{2} - \frac{q_\rho a}{2}) \bigg\} \, .
\label{a39} 
\eneqa
As we can see from Eq.~(\ref{a38}) and (\ref{a39}) the vertices are already order $g_0$ and $g_0^2$, so 
pluging into Eq~(\ref{a25}) is already more than $O(g_0^2)$. This imply that the Faddeev-Popov terms 
do not contribute to order $g_0^2$.

Concerning the vertices of $S_\mu$ for a 1-loop calculation we need the vertices corresponding 
to one-gluon-one-gluino, the two-gluons-one-gluino and finally the three-gluons-one gluino. 
They can be calculated from (\ref{a15}). The vertex one-gluon-one-gluino 
(using Eq.~(\ref{a53})) is
\beeq
\tilde{S}_{1 \mu, \rho \tau}^{abc}(q,p) = - \frac{2 i}{a} \delta_{a b} \sigma_{\rho \tau} \gamma_\mu 
\cos(\frac{p_\tau a}{2}) \sin(p_\rho a) \, ,
\eneq
which reduce to the continuum one in the limit $a \to 0$ (see Eq.~(\ref{a54})), 
while the vertex two-gluons-one-gluino is
\beeqa
\tilde{S}_{2 \mu, \rho \tau}^{abc}(q,p_1,p_2) &= & \frac{1}{2} g_0 f_{abc} \bigg\{ \sigma_{\rho \tau} \gamma_\mu 
\bigg[ \nonumber \\
&& \hspace{-2 cm} \cos(\frac{p_{1 \rho}a}{2} + \frac{p_{2 \rho}a}{2}) \cos(\frac{p_{1 \rho}a}{2}) 
\cos(\frac{p_{1 \tau}a}{2} + p_{2 \tau}a) + \nonumber \\  
&& \hspace{-2 cm} \cos(\frac{p_{1 \tau}a}{2} + \frac{p_{2 \tau}a}{2}) \cos(\frac{p_{2 \tau}a}{2}) 
       \cos(\frac{p_{2 \rho}a}{2} + p_{1 \rho}a) - \nonumber \\ 
&& \hspace{-2 cm} \sin(\frac{p_{1 \rho}a}{2} + \frac{p_{2 \rho}a}{2}) \sin(\frac{p_{1 \rho}a}{2}) 
               \cos(\frac{p_{1 \tau}a}{2}) - \nonumber \\
&& \hspace{-2 cm} \sin(\frac{p_{1 \tau}a}{2} + \frac{p_{2 \tau}a}{2}) \sin(\frac{p_{2 \tau}a}{2}) 
       \cos(\frac{p_{2 \rho}a}{2}) \bigg] - \nonumber \\ 
&& \hspace{-2 cm} \delta_{\rho \tau} \sum_\alpha  \sigma_{\alpha \tau} \gamma_\mu 
 \sin(\frac{p_{1 \tau}a}{2} + \frac{p_{2 \tau}a}{2}) \bigg(\sin(p_{1 \alpha}a) - \nonumber \\ 
&& \hspace{2 cm} \sin(p_{2 \alpha}a) \bigg) \bigg\} \, .
\eneqa
We do not presented here the three-gluons-one-gluino vertex because its contribution 
to the last Feynman diagram for the supercurrent, in Fig.~\ref{Fig1}, is zero by color considerations.

\section{Off-shell regime}
\label{appendixOFFSHELL}
\begin{figure}
\begin{center}
\epsfig{file=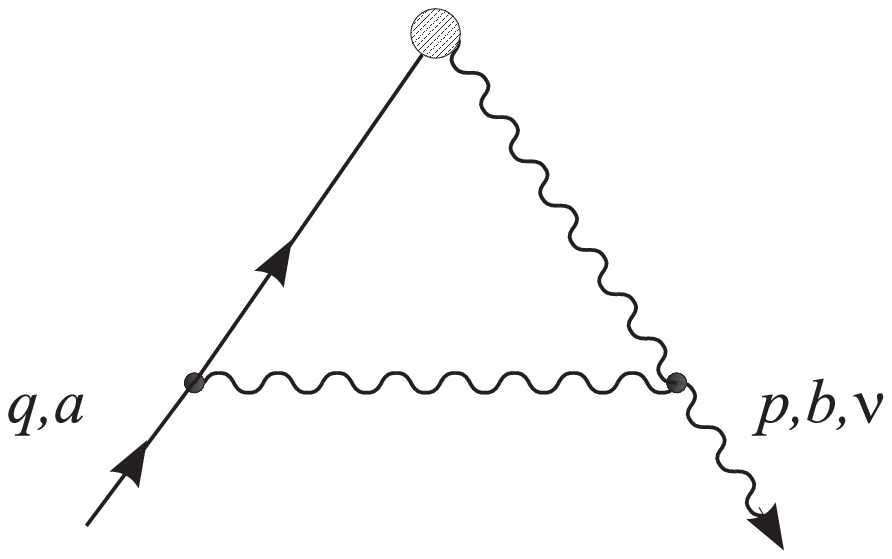,width=0.2\textwidth,height = 0.2\textwidth}
\epsfig{file=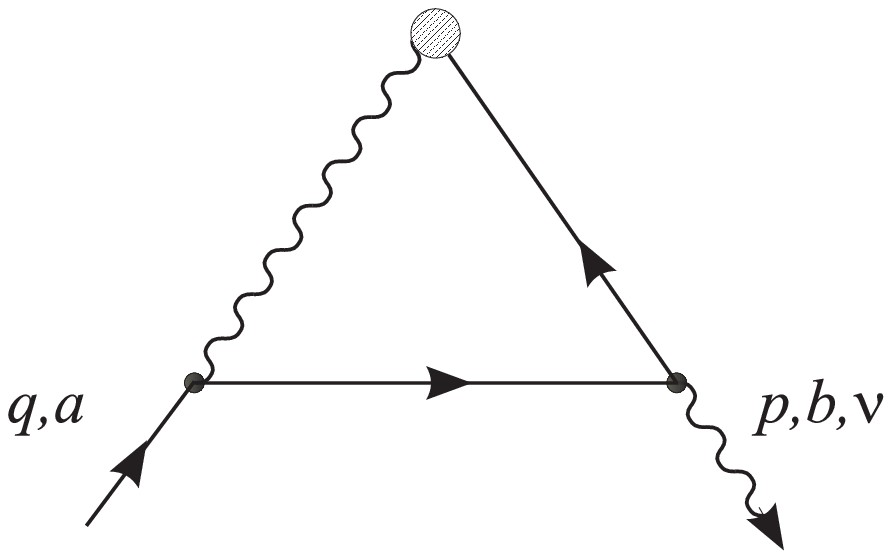,width=0.2\textwidth,height = 0.2\textwidth}
\epsfig{file=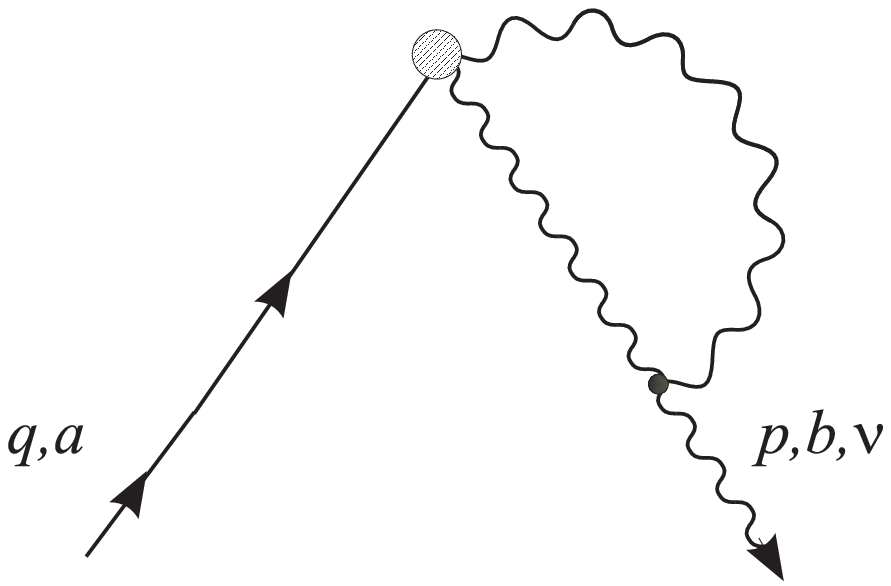,width=0.2\textwidth,height = 0.2\textwidth}
\epsfig{file=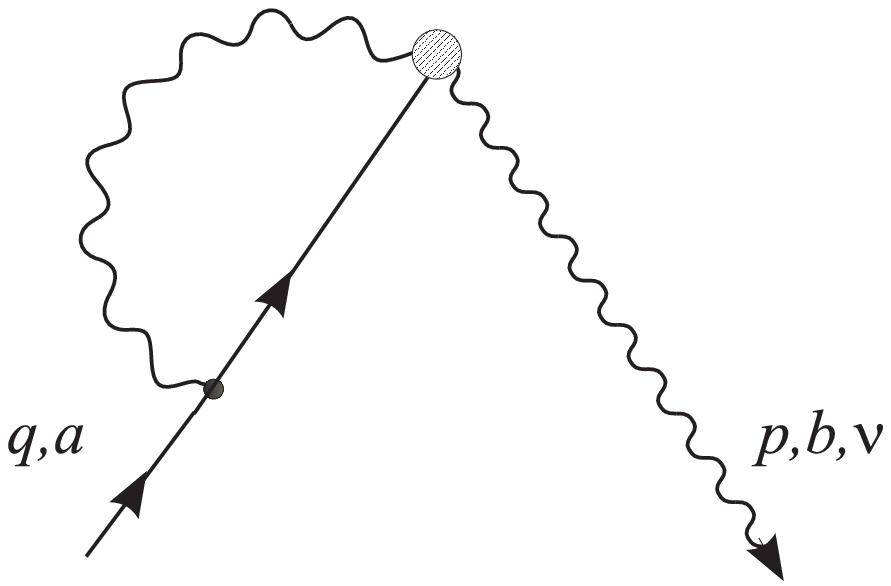,width=0.2\textwidth,height = 0.2\textwidth}
\epsfig{file=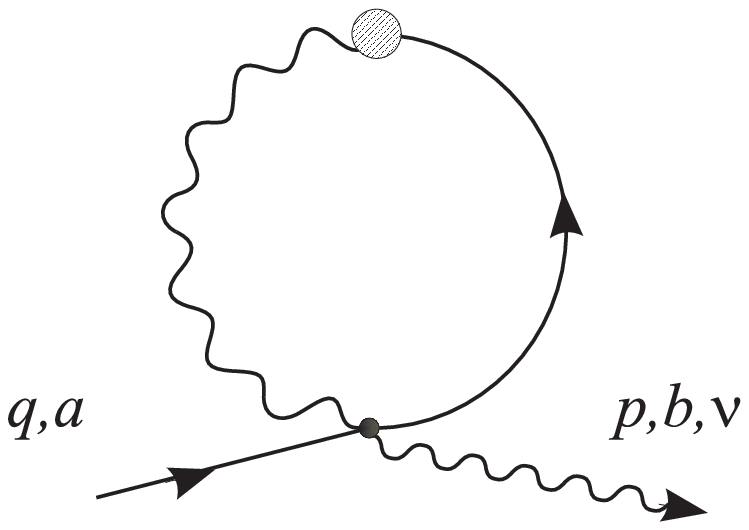,width=0.2\textwidth,height = 0.2\textwidth}
\epsfig{file=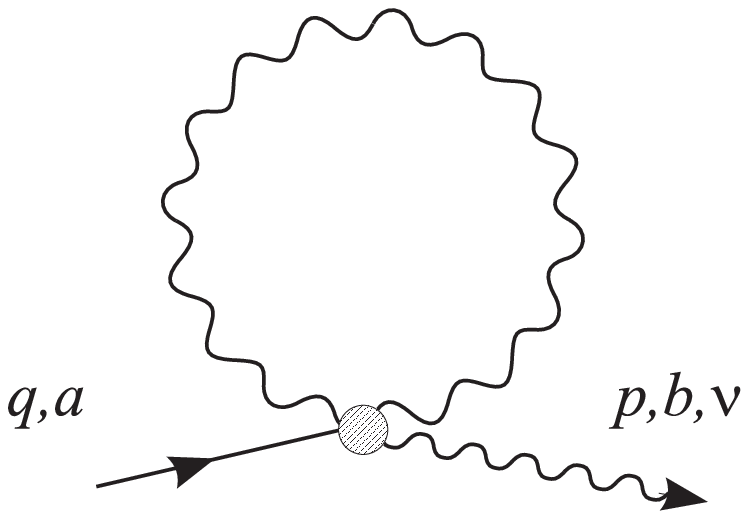,width=0.2\textwidth,height = 0.2\textwidth}
\caption{\sl Diagrams contributing for the supercurrent and the gauge fixing term. The grey blob correspond to the operator insertion in which flows a 
momentum $(p-q)$.} 
\label{Fig1}
\end{center}
\end{figure}

\begin{figure}
\begin{center}
\epsfig{file=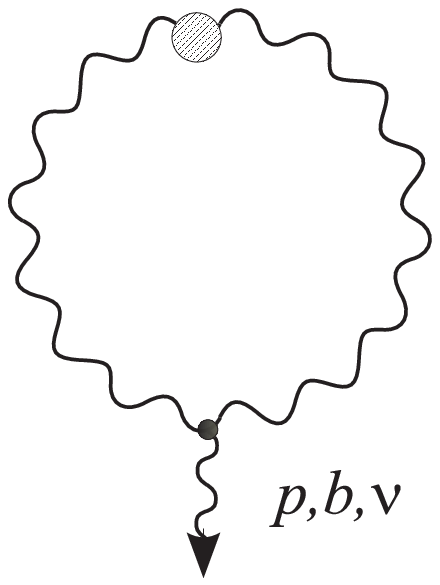,width=0.2\textwidth,height = 0.2\textwidth}
\epsfig{file=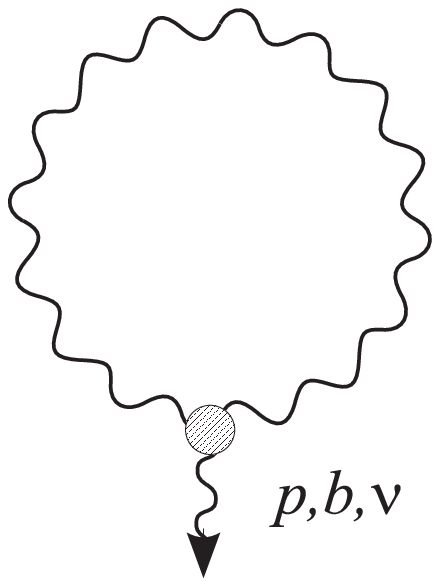,width=0.2\textwidth,height = 0.2\textwidth}
\caption{\sl Non-zero diagrams contributing to the contact terms.} 
\label{Fig2}
\end{center}
\end{figure}

\noindent
In order to separate the contribution of $T_\mu$ and $S_\mu$ at tree-level,
we can not impose $p =q$, that would greatly simplify the calculation. We are forced to  
use general external momenta $p$ and $q$ (while the momentum transfer of the operator insertion 
is $(p-q)\neq 0$, see Fig.~\ref{Fig1}).
Once the external momenta has been extracted from the propagators, in order to get the value of 
$Z_T$, the semplification $p^2 = q^2$ and $p \cdot q = 0$ is used. This is still an 
off-shell regime which simplify the dilogarithm functions. 

At 1-loop order, two propagators integrals are tabulated in \cite{kawai,martinelli}
while three propagator integrals on the lattice are tabulated in
\cite{haris} in terms of lattice constants plus the following continuum conterparts
\beeq
I_{0; 1 \mu; 2 \mu \nu; 3 \mu \nu \rho}(p,q) =  \frac{1}{\pi^2} \int d^4k 
\frac{1; k_\mu; k_\mu k_\nu; k_\mu k_\nu k_\rho}{k^2 (k+p)^2 (k+q)^2} \, . 
\eneq
With the help of \cite{velt,ball} one can give the expression for $I_0(p,q)$ and write down 
recursively $I_{1 \mu}(p,q)$, $I_{2 \mu \nu}(p,q)$, $I_{3 \mu \nu \rho}(p,q)$ in terms of 
the scalar functions $p^2$, $q^2$, $p \cdot q$ and $I_0$, plus Lorentz structures.
As an example \cite{ball}: $I_0(p,q)$ is a complicated function of 
$p$ and $q$, in terms of the dilogarithm as follows 
\beeqa
I_0(p,q) &=&  \frac{1}{\Delta} \bigg[ \mbox{Li}_2 \bigg( \frac{p \cdot q - \Delta}{q^2} \bigg) -
 \mbox{Li}_2 \bigg( \frac{p \cdot q + \Delta}{q^2} \bigg) + \nonumber \\ 
&& \frac{1}{2} \mbox{Ln}\bigg( \frac{p \cdot q - \Delta}{p \cdot q + \Delta} \bigg)
  \mbox{Ln} \bigg( \frac{(q - p)^2}{q^2} \bigg) \bigg] \, , 
\label{a19}
\eneqa
where $\Delta$ is the triangle function defined as 
\beeq
\Delta^2 = (p \cdot q)^2 - p^2 q^2
\label{a17}
\eneq
and 
\beeq
\mbox{Li}_2(x) = - \int^x_1 \frac{\mbox{Ln}\, t}{t - 1} dt
\eneq
is the dilogarithm.

Following the reference \cite{ball} where a tensor decomposition of 
$I_{1 \mu}(p,q)$, $I_{2 \mu \nu}(p,q)$, $I_{3 \mu \nu \rho}(p,q)$ is used, it is shown that 
all the integrals can be written in terms of $I_0$ and others scalars,
\beeq
I_{1 \mu}= I_1(p,q) p_\mu + I_1(q,p)q_\mu \, , 
\eneq
where
\beeqa
I_1(p,q)&=& \frac{1}{\Delta^2} \bigg[ q^2 \mbox{Ln}\bigg(\frac{(q-p)^2}{q^2}\bigg) - 
p \cdot q \mbox{Ln}\bigg(\frac{(q-p)^2}{p^2}\bigg) + \nonumber \\ 
&& \hspace{1 cm} \frac{q^2 p \cdot(q-p)}{2}I_0 \bigg] \, .
\label{a40}
\eneqa

The integral $I_{2 \mu \nu}$ is symmetric in $\mu$ and $\nu$ as well as under $p \leftrightarrow q$
and hence has the following tensor decomposition
\beeqa
I_{2 \mu \nu}&=& \delta_{\mu \nu} I_A + \bigg(p_\mu p_\nu - \frac{\delta_{\mu \nu}}{4} p^2 \bigg) 
I_B(p,q) + \nonumber \\ 
&& \bigg( p_\mu q_\nu + q_\mu p_\nu - \frac{\delta_{\mu\nu}}{2} p \cdot q \bigg) I_C + \nonumber \\
&& \bigg( q_\mu q_\nu - \frac{\delta_{\mu \nu}}{4} q^2 \bigg) I_B(q,p) \, ,
\label{a41}
\eneqa
where $I_A$, $I_B$ and $I_C$ are symmetric under $p \leftrightarrow q$ and tabulated in \cite{ball}.
In this reference an explicit expression for $I_{3 \mu \nu \lambda}$ is presented, which is quite 
complicated and we do not reported here.

The general result for arbitrary $p$ and $q$ using (\ref{a19}),
(\ref{a40}), (\ref{a41}) and the corresponding expression for $I_{3 \mu \nu \lambda}$ 
(in \cite{ball})
contains a huge quantities or terms (sometimes up to 1000 terms). 
Therefore a semplification which still leave us in the off-shell regime is required.
Let us rewrite (\ref{a17}) in the following way
\beeq
\Delta^2 = - p^2 q^2 \bigg( -\frac{(p \cdot q)^2}{p^2 q^2} + 1 \bigg) \, ,
\label{a42}
\eneq
where $p \cdot q = p \, q \cos\alpha$, where $ 0 < \alpha < \pi$. 
This imply that $(p - q)^2 = p^2 + q^2 - 2 p \, q \cos \alpha$.

By using Eq.~(\ref{a42}) it is possible to simplify $I_0$, $I_1$ $I_2$ and $I_3$. 
In fact, 
\beeqa
\Delta &=& i \sqrt{p^2 q^2 \bigg( 1 - \frac{(p \cdot q)^2}{p^2 q^2} \bigg)} \nonumber \\
     & =& i \sqrt{p^2} \, \sqrt{q^2} \sqrt{1 - \cos \alpha^2} \, .
\label{a18}
\eneqa
Substituting (\ref{a18}) in (\ref{a19}) we have
\beeqa
I_0 &= & \frac{1}{i p \, q \sqrt{ 1 - \cos \alpha^2} } \Bigg[ \nonumber \\
&& \mbox{Li}_2 \bigg( \frac{p \, q \cos \alpha - i p \, q \sqrt{1 - \cos \alpha^2}}{q^2} 
\bigg) - \nonumber \\  
&& \mbox{Li}_2 \bigg( \frac{p \, q \cos \alpha + i p \, q \sqrt{1 - \cos \alpha^2}}{q^2} \bigg) +
\nonumber \\
&& \hspace{-0.5 cm} \frac{1}{2} \mbox{Ln} \bigg( \frac{p \, q \cos \alpha - i p \, q \sqrt{1 - \cos \alpha^2}}
{p \, q \cos \alpha + i p \, q \sqrt{1 - \cos \alpha^2}}\bigg) \times \nonumber \\
&& \hspace{0.5 cm} \times \mbox{Ln} \bigg( \frac{p^2 + q^2 - 2 p \, q \cos \alpha}{q^2} \bigg) \Bigg] \, ,
\eneqa
simplifying we have
\beeqa
I_0 &= & \frac{1}{i p \, q \sin \alpha } \bigg[ \mbox{Li}_2 
\bigg( \frac{p}{q} (\cos \alpha - i \sin \alpha ) \bigg) - \nonumber \\
&& \hspace{1 cm} \mbox{Li}_2 \bigg( \frac{p}{q} ( \cos \alpha + i \sin \alpha ) \bigg) + \nonumber \\
&& \hspace{-1.8 cm} \frac{1}{2} \mbox{Ln} \bigg( \frac{\cos \alpha - i \sin \alpha}{\cos \alpha + 
i \sin \alpha} \bigg) \mbox{Ln} \bigg( \frac{p^2}{q^2} + 1 - \frac{2 p}{q} \cos \alpha \bigg) \bigg]
\eneqa
and finally 
\beeqa
\hspace{-1.8 cm} I_0 &= & 
\frac{1}{i p \, q \sin \alpha} \Bigg[\bigg( \mbox{Li}_2(\frac{p}{q} 
\exp^{-i \alpha}) - \mbox{Li}_2(\frac{p}{q} \exp^{i \alpha}) \bigg) \nonumber \\ 
&& \hspace{-0.6 cm} + \frac{1}{2} \mbox{Ln}\bigg( \exp^{-2 i \alpha} \bigg) \mbox{Ln} \bigg(\frac{p^2}{q^2} +
1 - \frac{2 p}{q} \cos \alpha \bigg) \Bigg] \, ,
\eneqa
or 
\beeqa
I_0 & = & \frac{1}{p \, q \sin \alpha} \bigg[ \frac{1}{i} \bigg( \mbox{Li}_2(\frac{p}{q} 
\exp^{-i \alpha}) - \mbox{Li}_2(\frac{p}{q} \exp^{i \alpha}) \bigg) - \nonumber \\
&& \alpha \mbox{Ln} \bigg(\frac{p^2}{q^2} + 1 - \frac{2 p}{q} \cos \alpha \bigg) \bigg] \, .
\label{a20}
\eneqa

Using Eq.~(\ref{a20}) we can now simplify the recursive expressions for $I_1$, $I_2$ and $I_3$.
Let us define
\beeq
\hat{p}_\mu \equiv \frac{p_\mu}{\sqrt{p^2}}
\eneq
where clearly $ |\hat p_\mu| = 1$. The semplification in Eq.~(\ref{a51}) correpond to
$\alpha = \frac{\pi}{2} $ (then, we have $\cos \alpha = 0$ and $\sin \alpha = 1$) and 
$p^2 = q^2 $, which correspond to the substitution $p \to q$ in all the results \footnote{Notice that
the substitution $p \to q$ is made {\em after} the external momenta from the diagrams has been extracted from the 
propagators, and the IR have been deal, using the Kawai procedure. The calculation of each diagram
has been done it in a completely off-shell regime but in order to read the values of the renormalization
constants, a proper semplification should be done it. Only at the very end of the calculation the substitution
$p \to q$ is made.}.

%\vspace{-5cm}
%\vspace{-5cm}

\end{document}